\documentclass[
superscriptaddress,
amsmath,
amssymb,
aps,
reprint,
pre,
twocolumn,
nofootinbib,
]{revtex4-2}
\usepackage{graphicx}
\usepackage{dcolumn}
\usepackage{bm}
\usepackage{color}
\usepackage[colorlinks]{hyperref}
\usepackage{xcolor}
\usepackage{soul}

\usepackage{dsfont}
\usepackage{pst-node}%
\usepackage{amsmath, mathtools,amssymb,booktabs}
\usepackage{algorithm}
\usepackage{algpseudocode}
\usepackage{physics}
\usepackage{array}
\usepackage{tabulary}
\usepackage[utf8]{inputenc}
\usepackage[T1]{fontenc}

\newcolumntype{K}[1]{>{\centering\arraybackslash}p{#1}}
\newcommand{\Gam}[2]{\textit{Gamma}\,\left(#1,\; #2\right)}

\newcommand{\appropto}{\mathrel{\vcenter{
  \offinterlineskip\halign{\hfil$##$\cr
    \propto\cr\noalign{\kern2pt}\sim\cr\noalign{\kern-2pt}}}}}
\begin{document}

\title[]{A minimal model of pan-immunity maintenance by horizontal gene transfer in the ecological dynamics of bacteria and phages}


\author{Wenping Cui}
\email{wenpingcui@kitp.ucsb.edu}
\affiliation{Kavli Institute for Theoretical Physics, Santa Barbara, California 93106, USA}

\author{Jemma M. Fendley}
\affiliation{Department of Physics, University of California Santa Barbara, Santa Barbara, California 93106, USA}

\author{Sriram Srikant}
\thanks{Deceased 1990-2025}
\affiliation{Department of Biology, Massachusetts Institute of Technology, Cambridge, Massachusetts 02139, USA}

\author{Boris Shraiman}
\affiliation{Kavli Institute for Theoretical Physics, Santa Barbara, California 93106, USA}
\affiliation{Department of Physics, University of California Santa Barbara, Santa Barbara, California 93106, USA}
\date{\today}


\begin{abstract}
Bacteria and phages have been in an ongoing arms race for billions of years. To resist phages bacteria have evolved numerous defense systems, which nevertheless are still overcome by counter-defense mechanisms of specific phages. These defense/counter-defense systems are a major element of microbial genetic diversity and have been demonstrated to propagate between strains by Horizontal Gene Transfer (HGT). It has been proposed that the totality of defense systems found in microbial communities collectively form a distributed "pan-immune" system with individual elements  moving between strains 
via ubiquitous HGT. Here, we formulate a Lotka-Volterra type model of a bacteria/phage community interacting via a combinatorial variety of defense/counter-defense systems and show that HGT enables stable maintenance of diverse defense/counter-defense genes in the microbial pan-genome even when individual microbial strains inevitably undergo extinction. This stability requires the HGT rate to be sufficiently high to ensure that some descendant of a "dying" strain survives, thanks to the immunity acquired through HGT from the community at large, thus establishing a new strain. 
This mechanism of persistence for the pan-immune gene pool is fundamentally similar to the "island migration" model of ecological diversity, with genes moving between genomes instead of species migrating between islands.

\end{abstract}

\maketitle

{\bf Introduction}
Bacterial viruses (phages) are the most abundant and diverse organisms on the planet. 
They exert substantial selection pressure on microbial communities by predation. For instance, it is estimated that marine viruses kill about 20\% of all ocean microbes each day \cite{suttle2007marine}. Facing phage predation, bacteria have developed different defense systems and strategies to interrupt the phage replication process, including restriction-modification (RM), abortive infection, and CRISPR–Cas systems \cite{barrangou2007crispr, doron2018systematic}. On the other side of this evolutionary conflict \cite{samson2013revenge}, phages have evolved counter-defense genes like the RM-inhibiting {\it ocr} \cite{atanasiu2001characterisation,  walkinshaw2002structure}, protein inhibitors like {\it dmd} or {\it tifA} against abortive infection toxin-antitoxin systems \cite{otsuka2012dmd, srikant2022evo}, and many different anti-CRISPR genes \cite{bondy2013bacteriophage, pawluk2018anti}. 

Phages and bacteria have been a useful model to study evolution in the lab since the beginning of molecular biology.  Although long-term co-existence can arise under conditions of genetic or phenotypic stratification \cite{koskella2014bacteria, pyenson2024diverse}, or in spatially heterogeneous environments \cite{gomez2011bacteria}, co-culturing experiments in the lab often result in either phages or bacteria fixing in the culture driving the other to extinction in the long term \cite{lenski1985constraints, borin2023rapid}. Further, mathematical models of such systems tend to show that an advantageous bacterium or phage would always dominate, with the ecosystem losing diversity in the long run  \cite{childs2014crispr, pilosof2020network}, unless given unrealistically high mutation rates \cite{xue2017coevolution, martis2022eco}. However, we know that the evolutionary conflict has persisted in nature for billions of years, leading to the phylogenetic diversity we see today. Understanding the persistent coexistence of diverse bacteria and phages must involve stochastic ecological dynamics coupled with evolution, in addition to the genetic systems that the two sides use against the other \cite{chevallereau2022interactions}.  

Recent studies have found that many defense systems are co-localized on bacterial genomes in so-called defense islands \cite{doron2018systematic, gao2020diverse}, and some are located on mobile genetic elements \cite{rousset2021prophage, vassallo2022functional}. It has also been noted that the phylogeny of orthologous defense genes is discordant with the phylogeny of the core genomes of the strains that harbor them \cite{bernheim2020pan,ding2018panx}, suggesting frequent horizontal transfer. The apparent mobility of defense systems is not surprising given that they are beneficial only in the presence of certain phages, while presenting a persistent cost, e.g. in metabolic resources of a cell. Given the diversity and abundance of defense systems, a given genome contains only a subset of existing defense genes \cite{hochhauser2023defense, rousset2021prophage, vassallo2022functional}. Evidence of frequent gain and loss of defense systems, along with the fact that in natural ecosystems the defense "arsenal" is distributed across strains, has motivated the \textit{pan-immunity} hypothesis, which suggests the view of the bacterial pan-genome as a collective "immune system", with the diversity of defense systems maintained by transient selection and frequent horizontal gene transfer  (HGT) \cite{kauffman2022resolving, legault2021temporal, piel2022phage,bernheim2020pan}. Phages are also limited in the number of counter-defense genes they contain due to the constraint on their packaged genome size, and therefore are limited in the number of bacterial strains they can successfully infect \cite{srikant2022evo}. Thus, phages must evolve by gaining different combinations of counter-defense genes to continue to persist in a bacterial community that is shuffling a per-genome immune profile \cite{legault2021temporal, piel2022phage, samson2013revenge}.

\begin{figure}
\centering
\includegraphics[width=0.49\textwidth]{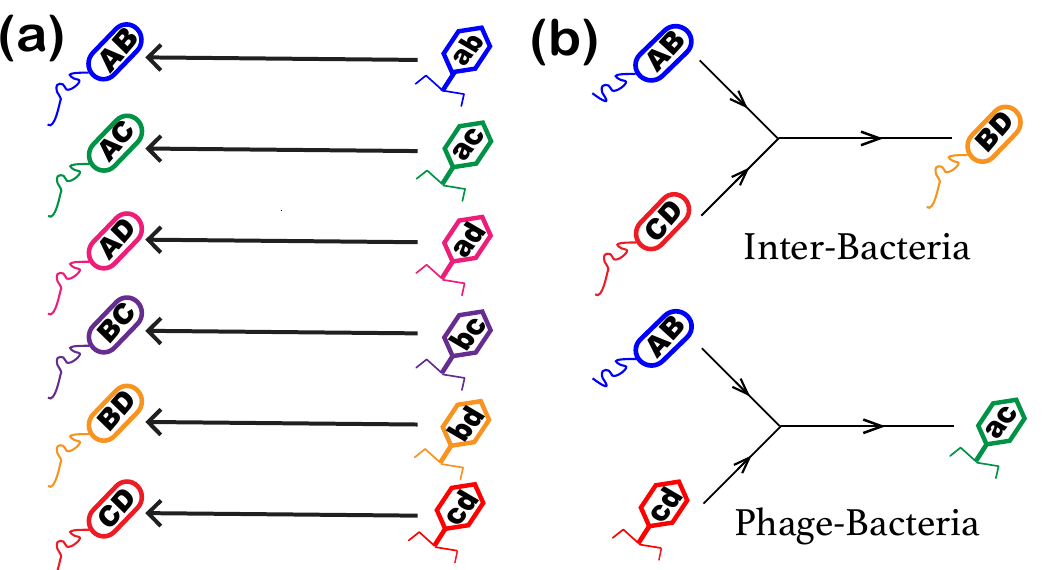}
\caption{
{\bf Model bacteria and phage genotypes and HGT processes.} {\bf (a)} Interactions between bacteria and phages for the case of $L=4$ different TA systems. The capital letter denotes a specific toxin in the bacterium; the lowercase letter denotes a corresponding antitoxin gene in the phage. Bacterium AB will carry the toxins A and B, in addition to the antitoxins a and b.  
{\bf (b)} Scheme for the new clones formed by inter-bacteria and phage-bacteria HGT processes. }
\label{fig:scheme}
\end{figure}

The pan-immunity hypothesis, while not yet empirically established, extends the concept of diversity into multi-level space \cite{childs2012multiscale}. Many of the existing models primarily focus on diversity at a single level, typically the number of surviving species \cite{bunin2017ecological, cui2020effect, biroli2018marginally, servan2018coexistence}. However, the pan-immunity hypothesis emphasizes the importance of considering the diversity of genes and genomes (strains)  – combinatorial sets of different genes – separately. Since the gene space is considerably smaller than that of genotypes, it becomes more achievable for diverse genes to persist over time, regardless of the rapid extinction and turnover of genomes.

Toxin-antitoxin (TA) systems are widely distributed across bacterial genomes, with diverse TA systems playing a role in phage defense \cite{leroux2022toxin}. In TA systems, the toxin activity, that is normally inhibited by the antitoxin, is triggered by phage infection, leading to a collapse of phage replication. Given the mechanistic diversity and abundance of these systems in bacterial genomes, phages must evolve to evade a wide variety of TA systems. One evolutionary strategy phages have used is to maintain antitoxins to inhibit TA systems and allow infection of bacteria \cite{otsuka2012dmd, srikant2022evo}. Phages have even been shown to acquire the antitoxin from the TA locus by HGT to evolve the ability to infect the bacteria \cite{Blower2013}. This is possible as bacterial immune systems do not provide absolute protection against phage infection. TA systems can therefore play a key role in phage-bacteria co-evolutionary dynamics providing an immune barrier to viral replication but also providing the basis of viral resistance by HGT.

In this article, we develop a stochastic Lotka-Volterra (LV) model that incorporates HGT between genotypes and explores the feasibility of pan-immunity. Our minimal model can exhibit typical LV behaviors: demographic-noise-driven extinction \cite{dobrinevski2012extinction} and persistent (oscillatory) coexistence, realized for different HGT rates. Surprisingly, between these two phases, we identify another regime where the pan-immunity hypothesis operates, and genes can persist along with the "boom-bust" dynamics of individual strains or particular genotypes. Inspired by recent theoretical progress in islands-migration models \cite{pearce2020stabilization}, we derive novel criteria for the gene and genotype persistence regimes. Our results suggest that a relatively small HGT rate, inversely proportional to the total population size, is sufficient to maintain the coexistence of diverse genes, independent of the combinatorial diversity of genotypes, explaining how the numerous defense and counter-defense systems can persist despite strong selections in nature \cite{hossain2021viral, hussain2021rapid}.

\noindent {\bf Model:}
We shall focus on bacteria-phage interactions assuming for simplicity that susceptibility to specific phage infection is the only heritable phenotype that affects relative fitness of bacterial species and similarly the ability to infect a specific bacterium is the only relevant distinction between phages. Thus, the relevant space of genotypes of bacteria and phages is defined respectively by the defense and counter-defense genes.

To illustrate the effect of HGT, it will suffice to frame our model in the context of the toxin/antitoxin (TA) paradigm of phage defense \cite{leroux2022toxin}. 
The key elements of TA systems can be abstracted in terms of a toxin gene, {\it A}, and its specific antitoxin, {\it a}. A bacterium carrying $A$ must also carry $a$ and, we shall assume, can only be infected by a phage carrying the antitoxin $a$. We shall posit the existence of a large number of distinct specific toxin/antitoxin pairs denoting them by $L$ different uppercase/lowercase letters with $L \gg 1$. We emphasize that on this level of abstraction the toxin/antitoxin paradigm captures the general aspects of a broad variety of actual defense/counter-defense systems, without delving into the biochemical and genetic complexity of their mechanisms. This will suffice for our goal of providing a mathematical underpinning for the "pan-immunity" hypothesis.

Each bacterium and phage can carry multiple defense and counter-defense genes respectively, introducing combinatorial complexity into ecological interactions.  We start with each bacterial and phage strain carrying just two distinct TA genes. Correspondingly, there are $K=L(L-1)/2$ possible genotype combinations in total.  Fig. \ref{fig:scheme}a shows an example of $L=4$, where there are $K=6$ possible bacteria-phage pairs.  

Let $B_{ij}$ and $V_{ij}$ denote the population sizes of bacterial and phage strains (number of individuals in the system), respectively, with index $ij$ specifying the genotype via the labels $i,j$ of its TA genes. 
By construction, $B_{ij}$ and $V_{ij}$ are symmetric matrices, and $i\neq j$.  The \textit{eco-evolutionary dynamics} of bacteria and phage populations may be described by the generalized LV equations:
\begin{subequations}\label{eq:sde_main}
  \begin{align}
\frac{dB_{ij}}{dt}&=sB_{ij} - \frac{\phi}{N_B} B_{ij} V_{ij}+\frac{r_B}{2N_B} \sum_{k,l}B_{ik}B_{lj},\\
\frac{dV_{ij}}{dt}&=  \frac{\beta\phi}{N_B} B_{ij} V_{ij} - \omega V_{ij} +\frac{r_V}{4 N_B} \sum_{k,l}\left( V_{ik}B_{lj} +V_{lj}B_{ik}\right), 
\end{align}
\end{subequations} 
 where $s$ is the per capita bacterial division rate,  $\beta$ is the burst size (dimensionless), and $\omega$ is the phage's  per capita death  rate. $\phi/N_B$ is the successful per capita infection and lysis rate so that the infection term does not scale with the number of individuals in the system, and $N_B$ is the total number of all bacterial individuals, which remains undetermined.    For simplicity, we neglect the potential dependence of these on TA genotype $ij$, effectively assuming that all TA systems have the same intrinsic fitness cost, so that each phage/bacterium pair follows similar dynamics.
 
 The final terms in eqs. \ref{eq:sde_main} represent the HGT process. As a minimal model - a simplified view of bacterial conjugation - we assume that a bacterium can replace one of its genes with a gene from another bacterium. Fig. \ref{fig:scheme}b top shows the bacterial recipient \textit{CD} randomly acquires TA gene \textit{B} from some  donor  \textit{AB} and transforms its genotype to \textit{BD}.  
Antitoxin genes can be passed on from one phage to another, via instances of co-infection of the same bacterium. Alternatively, they can be passed directly from a resistant bacterium to a phage genome \cite{srikant2022evo}. This is possible since bacterial immune systems are not perfect barriers to viral infection \cite{Blower2013}.
 Fig. \ref{fig:scheme}b (bottom) shows the latter. We are interested in the total number of transfers into a given strain per unit time, which is also measurable by statistical inference from experimental data \cite{liu2024dynamics},  rather than the rate per unit volume, as in mass action. Therefore, we define bacteria and phages to have per capita HGT rates $r_B$ and $r_V$, with respective terms shown in eqs. \ref{eq:sde_main}.

\begin{figure}
    \centering
\includegraphics[width=0.9\linewidth]{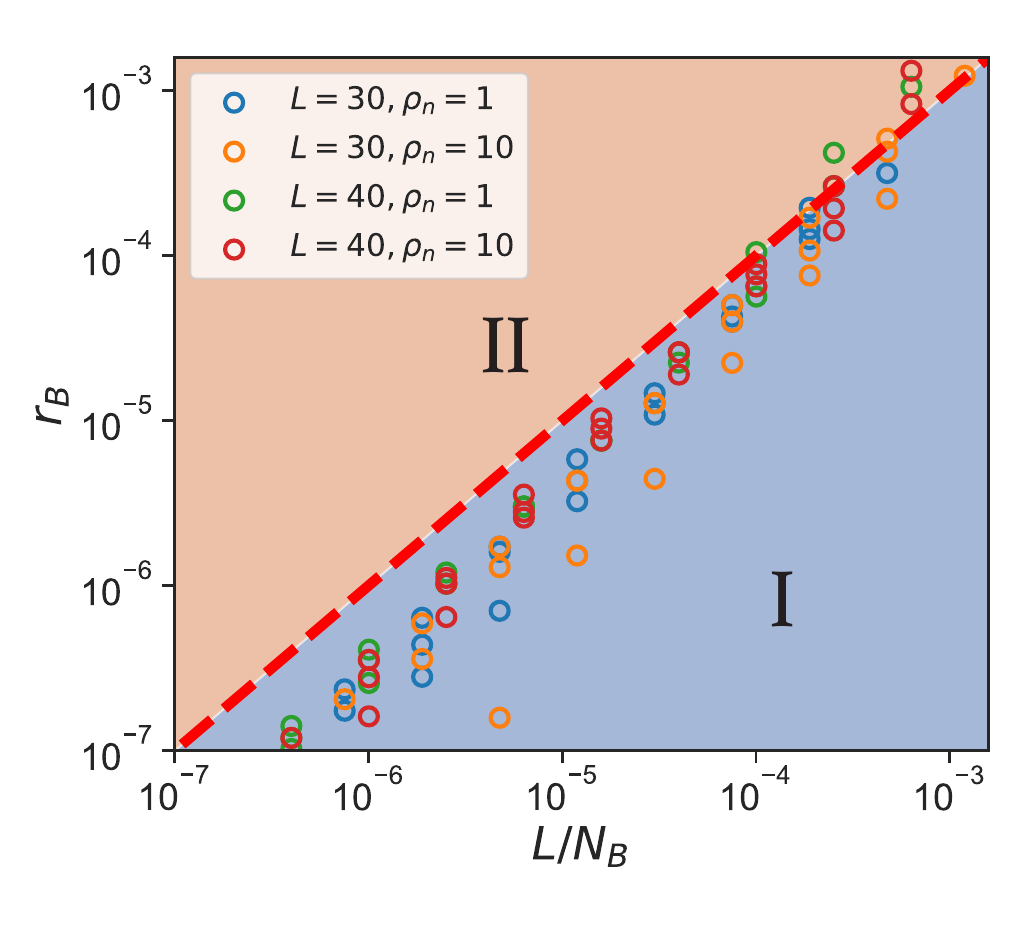}
    \caption{"Phase diagram" for unequal inter-bacteria and bacteria-phage HGT rates. The scatter points  indicate the numerically determined transition points between Regions I and II. We varied $N_B$ from $10^5$ to $10^8$ with $\rho_n= N_V/N_B=1, 10$ and $r_V = 100/N_B, 200/N_B, 400/N_B$, ensuring the system stays in Regime II. The red dashed horizontal line: $y=x$ is our theoretical estimate for the I/II boundary given by eq. \ref{eq:HGT_ratio_inequality}.}
    \label{fig:hostHGT}
\end{figure}

To investigate extinction driven by demographic noise, we generalize eqs. \ref{eq:sde_main} into a Poisson process framework. Specifically, bacteria and phages now follow stochastic growth with fitness defined by $s - \phi V_{ij}/N_B$ and $\beta \phi B_{ij}/N_B - \omega$, respectively. As the number of HGT events is proportional to the population size, it is convenient to reparameterize the bacterial and viral "fitness" terms, respectively, as $s(1 -  V_{ij}/n_V^*)$ and $\omega(B_{ij}/n_B^*-1)$, where $n_V^*=sN_B/\phi$ and $n_B^*=\omega N_B/(\beta \phi)$ are the phage and bacteria characteristic population sizes for each strain. 
For scaling purposes, we introduce the ratios $\rho_n = n_V^*/n^*_B$, $\rho_s = \omega/s$, and $\rho_r=r_V/r_B$ and rewrite $n_B^*=n^*_G$, and $r_B = r$. The two-species LV dynamics can always be rescaled to the anti-symmetric form \cite{goel1971volterra, pearce2020stabilization}. To further simplify our analysis, we initially investigate the perfectly anti-symmetric form by setting $\rho_s=\rho_n=\rho_r=1$. The summary of notations can be found in Supplementary Information (SI) Sec. \ref{sec:summaryvariables}. In SI Sec. \ref{sec:distinct_param} we show that our primary findings still hold in the more general parameter setting, and the case $\rho_r\neq 1 $ is considered later in Fig. \ref{fig:hostHGT}.

However, the total population sizes $N_B = \frac{1}{2}\sum_{i\neq j} B_{ij}$, and $N_V = \frac{1}{2}\sum_{i\neq j} V _{ij}$ are still undetermined and depend on complex relationships involving all system parameters.  To avoid solving sophisticated self-consistency relations for $N_B$ and $N_V$, we impose a hard constraint on the total population sizes. First, we initialize $s$, $\omega$, $N_B$ and $N_V$, and then taking $n_B^*=N_B/K$ and $n_V^*=N_V/K$. The population size constraint is achieved by introducing regulators which tune $s$ and $\omega$. Notably, we find that the collective phenomena associated with gene transfer are insensitive to $s$ and $\omega$ as they either cancel out or introduce order-1 corrections in most cases. The full description of our simulation can be found in SI Sec. \ref{sec:simulation}.

Biologically, the fixed total population size serves as a universal carrying capacity across strains. 
Our study focuses on strains competing within the same ecological niche, distinguished solely by their phage defense systems. Consequently, we adopt LV dynamics with fixed total population sizes, neglecting strain-specific carrying capacity terms in conventional consumer-resource models. Without phages, our model becomes the Wright-Fisher model \cite{crow2017introduction}; bacterial strains undergo random drift because of the finite population size, one strain dominates, and the ecosystem completely loses genetic diversity.

\begin{figure*}
\centering
\includegraphics[width=0.95\textwidth]{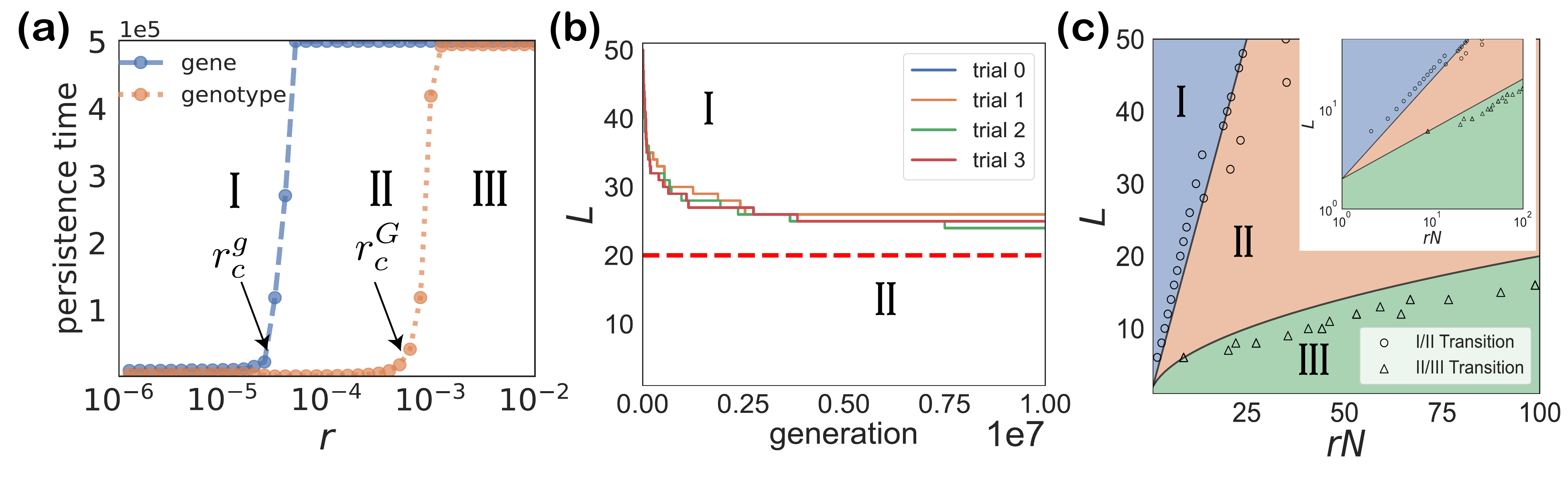}
\caption{
{\bf Different regimes of HGT-driven bacteria/phage dynamics.} {\bf (a)} Persistence time for genes and genotypes under different HGT rate $r$ (and fixed $N=10^6$ and initial $L=40$). Persistence time is defined by the simulation time (averaged over 5 trials) until the first gene or genotype is lost (with the upper limit at generation $T=5\times10^5$ imposed by the maximum simulation length). 
{\bf (b)} Genetic diversity decreases when its initial value exceeds $L_{max}$ (for $N=10^6$ and $r=2\times10^{-5}$). The solid lines are from 4 trials of simulations. The red dashed line is $L_{max}\propto rN$ given by our theory. 
{\bf (c)} Parameter regimes corresponding to three qualitatively different behaviors: I: unstable coexistence; II: persistence of genes with continuous turnover of genotypes; III: stable coexistence; each circle and triangle marker represents one-trial simulations to search for the transition boundary between I/II and  II/III at given parameters. The inset shows the same data but on a log-log scale. The black solid lines separating different regimes are our theoretical predictions given by eq. \ref{eq:critical_rs_main}: $L_{max}\propto rN$ (upper) and $K_{max}\propto rN$ (lower), respectively. The prefactors are estimated empirically.
Our default simulation parameters are $N=10^6$, $L=40$, $K=780$, and $s=5\times 10^{-3}$ unless specified otherwise.
}
\label{fig:phase}
\end{figure*}



\noindent {\bf Different regimes in the dynamics of genes and genotypes:} We would like to understand how HGT processes affect the distributions of bacteria and phage genotypes and the frequencies of specific TA genes in the pan-genome. Fig. \ref{fig:phase}a shows how long the system keeps its full gene and genotype diversity when all genotypes start from the mean population size $N/K$, with $N=N_B=N_V$.  For simplicity, we set $r= r_B=r_V$, as described in the previous section. And numerical simulations reveal three distinct regimes as a function of the HGT rate $r$, and thus two critical HGT rates $r_c^g$ and $r_c^G$ are defined with $r_c^g<r_c^G$. 

\textit{Regime I} is the {\it unstable-coexistence} regime which occurs for low HGT rate  $r<r_c^g$. In this regime any given strain (i.e. a particular $ij$ genotype) persists for a short time before going extinct and is unlikely to reappear via HGT. Loss of genotypes leads to a loss of genes (see Fig. \ref{fig:phase}a), reducing genetic diversity. The bacteria/phage system keeps losing genetic diversity until the "birth rate" of genotypes via HGT is sufficient to compensate the loss of genotypes under selection. The genetic diversity will then be sustained, but with a smaller number of genes (i.e. TA systems), $L_{max}$ (see Fig. \ref{fig:phase}b), and a larger average population size of different genotypes. Thus, the unstable-coexistence regime appears transiently while $L>L_{max}$, as shown in Fig. \ref{fig:phase}b, with the dependence of $L_{max}$ on $rN$ defining the boundary of the unstable coexistence regime in the phase diagram Fig. \ref{fig:phase}c. 

\textit{Regime II} is the regime of {\it genotype turnover} and gene persistence, which occurs in the intermediate range of HGT rate $r_c^g<r<r_c^G$. In this regime bacteria-phage pairs undergo \textit{boom-bust} behavior: a newly established bacterial genotype, formed by HGT, grows rapidly until infected by a phage which subsequently leads to a "crash", and then goes extinct. 

\begin{figure*}
\centering
\includegraphics[width=0.88\textwidth]{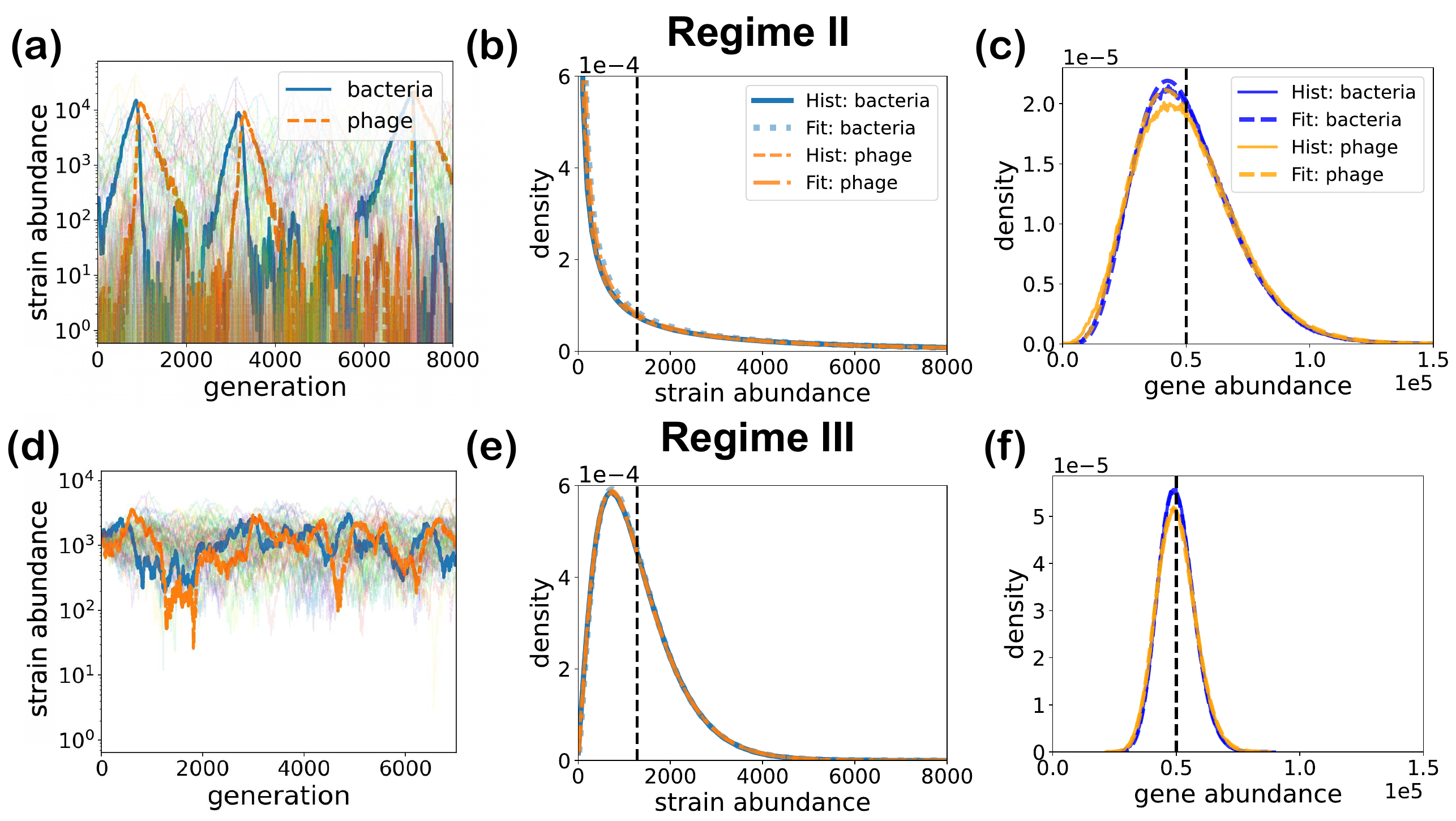}
\caption{
 {\bf Phage/bacteria population dynamics and the distributions of genes and genotypes in Regimes II and III}. {\bf  (a, d)} Time course  of multiple phage (orange dashed lines) and bacteria (blue solid lines) populations. One phage/bacteria-pair is emphasized to illustrate distinct dynamics in Regime II ($r=5\times10^{-5}$) and Regime III  ($r=10^{-3}$) in {\bf (a)} and {\bf (d)}  respectively. Corresponding
genotype {\bf (b, e)} and gene {\bf (c, f)} abundance distributions. Dotted and dash-dotted lines are the fitted Gamma distributions; the lines overlap so extensively that they are difficult to distinguish from one another. The vertical lines are the mean clone size and the mean population size for genes. }
\label{fig:gamma_distribution}
\end{figure*}

Critically, however, in this regime the HGT rate is high enough for one or more descendants of the clonal population to acquire immunity against the phage, via a horizontally-transferred TA gene. A successful transfer event establishes a new immune genotype, which will subsequently undergo a boom-bust cycle of its own. Fig. \ref{fig:gamma_distribution}a shows that the genotypes have a short lifetime, but HGT enables
TA genes to persist in the pan-genome of the community, effectively "surfing" from one boom-bust wave to another. This regime provides a model for the pan-immunity hypothesis. Only a fraction of all possible genotypes exist at any instant of time; however, the system can still maintain genetic diversity by distributing genes across the community.      

\textit{Regime III} is the regime of {\it genotype persistence} realized at a sufficiently high HGT rate $r>r_c^G$. Fig. \ref{fig:gamma_distribution}d shows the bacteria-phage pair undergoing \textit{stochastic oscillations} around a well defined mean population size without going extinct. In this regime frequent HGT spreads bacterial and phage genotypes over all available genotypes, making immune escape impossible, while also eliminating large boom-bust events. This is the endemic infection regime, which is also the regime of stable bacteria-phage coexistence.

Our simulations further demonstrate that the critical HGT rates $r_c^g$ and $r_c^G$ most sensitively depend on the total population size $N$ and the genetic diversity $L$. Consequently, we can alternatively express the critical curves for a given $N$ and $r$ in terms of $L_{max}$, representing the maximum genetic diversity and analogous to the carrying capacity in ecology. If the initial value of $L$ is larger than  $L_{max}$, the system starts at Regime I, which is unstable, and subsequently slides toward the edge of Regime II, as illustrated in Fig. \ref{fig:phase}b. Likewise, we also have the maximum genotypic diversity $K_{max}$, conditioned upon the coexistence of all possible genotypes, lying at the boundary between Regime II and Regime III. Fig. \ref{fig:phase}c provides the full phase diagram (we transform $K_{max}$ back to its corresponding $L_{max}$ for comparison in the same diagram), and we will give our theoretical analysis of the critical curves in the following sections. 

\noindent {\bf Phage/bacteria "ensemble":} 
What is the critical value of $r$ for the transition between regimes, given model parameters?
To identify suitable observable quantities to describe the highly non-trivial population dynamics of many phage-bacteria pairs in Fig. \ref{fig:gamma_distribution}ad, we shall follow \citeauthor{goel1971volterra}, as our system in eqs. (\ref{eq:sde_main}) falls into a broad class of anti-symmetric LV models investigated before \cite{goel1971volterra, pearce2020stabilization}. It is useful to define the Lyapunov function \cite{goel1971volterra} for a single bacteria-phage pair as
\begin{equation}\label{eq:Lapu0_main}
 E(B,V) =(B -n^*_G\log{\tfrac{B}{n^*_G}}) +(V-n^*_G\log{\tfrac{V}{n^*_G}}),
\end{equation}
which characterizes the distance between the current state and the steady state. In the deterministic limit $N\rightarrow \infty$, $E$ is determined by the initial condition and does not change with time, corresponding to a bacteria/pathogen population undergoing a neutrally stable periodic oscillation.

In the spirit of statistical mechanics, one may interpret $E$  as the energy of a "particle", and describe the state of the stochastic LV system with a large number of interacting strains by a canonical ensemble \cite{kerner1957statistical, goel1971volterra}.  In our case HGT processes play a role similar to particle collisions for energy exchange. 
 Then the probability density function (PDF) of microstates follows Boltzmann statistics which give the \textit{Gamma} distribution (see SI Sec. \ref{SI:canonical_ensemble} for details):
\begin{equation}\label{eq:gamma_main}
 n_G \sim \Gamma ({\tfrac{N}{K}},{\Theta}) = \frac{n_G^{\tfrac{N}{K\Theta} - 1} e^{-n_G/\Theta}}{\Theta ^{\tfrac{N}{K\Theta}}\Gamma (\tfrac{N}{K\Theta})}.
\end{equation}
For simplicity, we use $n_G$ to represent either bacteria or phage abundances since their 
distributions have the same general form. Its mean $n^*_G$ is given by $N/K$, and $\Theta$ is an unknown effective temperature for genotypes, characterizing the fluctuations of genotype abundances.  

The gene abundance $n_g$ is the sum of population sizes of all existing genotypes containing a specific gene $g$. We assume that the Gamma distribution ansatz can also be applied to the gene abundance:
 $n_g  \sim  \Gamma({2N/L},{\theta})$,
where $2N/L$ is the mean gene abundance (the factor 2 arising from two-gene genotypes), and $\theta$ is another unknown effective temperature for genes. 

Fig. \ref{fig:gamma_distribution} shows that Gamma distributions provide an excellent fit to the genotype and gene abundance distributions across Regime II and III, even though their population dynamics behave quite differently.

\noindent {\bf "Effective temperatures" for genes and genotypes:} 
We next address the relationship between the effective temperatures, $\theta$ and $\Theta$, that describe phage/bacteria statistics within the canonical ensemble framework, with the parameters controlling phage/bacteria dynamics. 

The HGT process can be approximated in the mean-field sense by treating the recombination terms as constant $rN/K$ source terms in both eqs. (\ref{eq:sde_main}).  This is because the sampling probability of a specific genotype from HGT is proportional to the product of marginal probabilities of genes it carries \cite{neher2011statistical}; at the genotype scale, the fluctuations of gene abundances can be averaged out with the law of large numbers (also see simulations in SI Fig. \ref{figS:gene_genotype_dynamics}), suggesting that the new clones can be uniformly sampled over the whole genotype space. Within this approximation, the effect of HGT on ecological dynamics is similar to the effect of "island migration", extensively studied in the context of LV models of ecological dynamics \cite{pearce2020stabilization}.

\begin{figure*}
\centering
\includegraphics[width=0.75\textwidth]{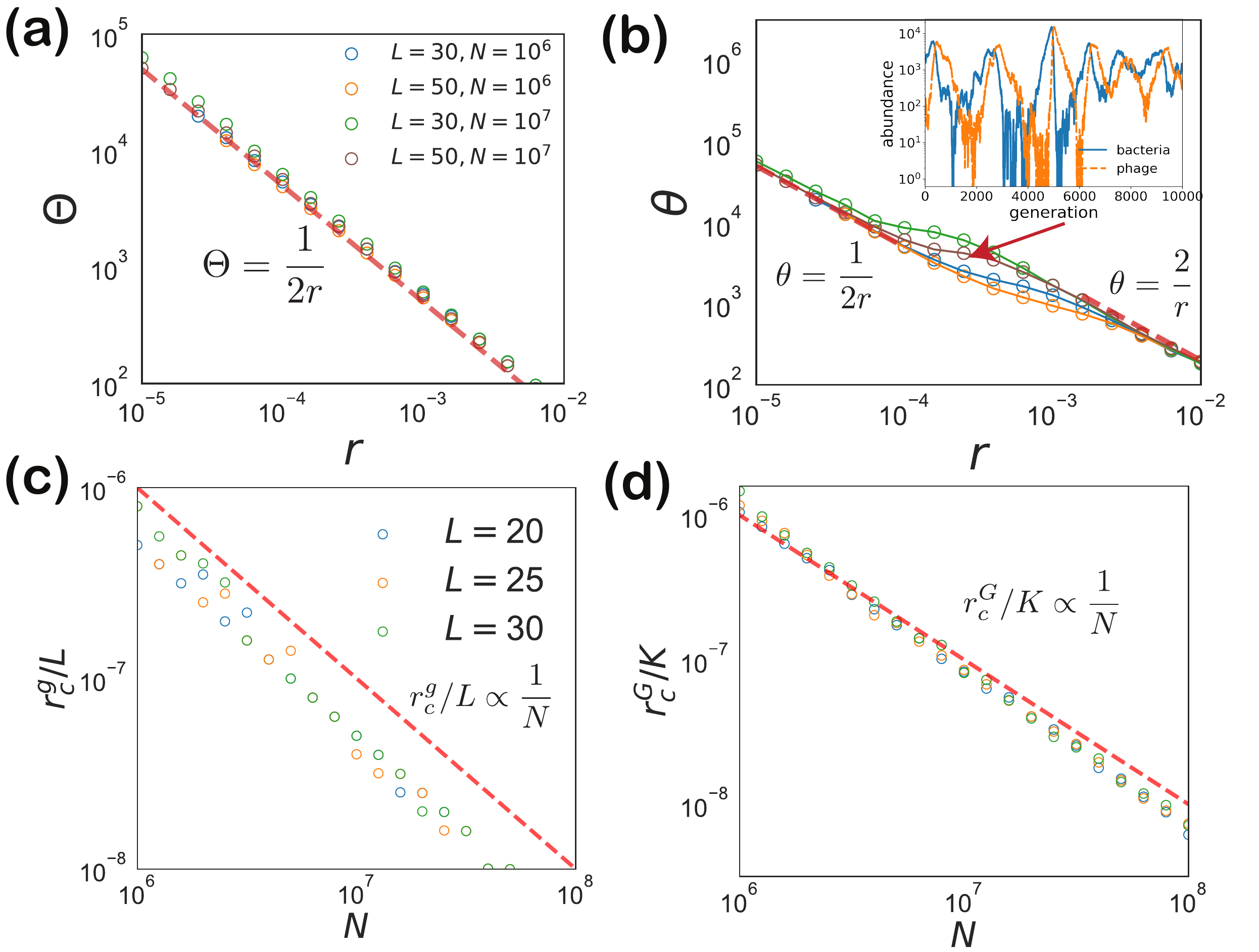}
\caption{ {\bf "Effective temperatures" and critical HGT rates.}
{\bf (a, b)} The scaling of $\Theta$ and $\theta$ in Gamma distributions of genotype and gene abundances at different HGT rates $r$. Red dashed lines are theoretical predictions. Inset in {\bf (b)} shows an example of a phage-bacteria pair exhibiting a mixture of boom-bust and stochastic-oscillation dynamics at the crossover between Regimes II and III. 
{\bf{(c, d)}} Critical (minimal) HGT rates for gene and genotype coexistence. The red dashed lines are our predictions. All data points are simulations with gene {\bf (c)} or genotype {\bf (d)} persistence time over $2\times 10^5$ generations.}
\label{fig:critical}
\end{figure*}

Using  It\^{o}'s lemma  \cite{gardiner1985handbook} and the mean-field approximations, we can write down the Lyapunov function dynamics for eqs. (\ref{eq:sde_main}) with with additional terms for demographic noise:
\begin{equation}
\begin{aligned}\label{eq:SDE_recombiantion_main}
dE = &\left(\tfrac{n^*_G}{2B} +\tfrac{n^*_G}{2V} -r\tfrac{N(n^*_G-B)}{KB} -r\tfrac{N(n^*_G-V)}{KV} \right) dt\\
& + \sqrt{\tfrac{(B-n^*_G)^2}{B} + \tfrac{(V-n^*_G)^2}{V}} d\eta.
\end{aligned}
\end{equation}
From the above equation, we see that demographic noise contributes a positive drift and drives $E$ to grow exponentially (see SI Sec. \ref{app:extinction}). 
The injection of new clones due to HGT processes contributes a balancing force to cancel the effect of demographic noise and stabilizes the system.

Without HGT, the demographic noise drives the system away from the steady state.  The oscillation amplitude grows until either the bacterium or the phage dies out and the coexistence becomes unstable after a typical persistence time proportional to the average clone size $N/K$ (see SI Sec. \ref{app:extinction} and Fig. \ref{figS:extinction}). Once any phage goes extinct first, its corresponding  bacterium becomes advantageous and drives the catastrophic extinction of other bacteria because of competitive exclusion.    

In Regime III where $r$ is sufficiently large, the system is stable and the average of $E$ does not change in thermal equilibrium. We then obtain $\Theta = \frac{1}{2r}$ by solving the self-consistency relation that the drift part averaged over the canonical ensemble is zero. With similar calculations for the mean-field gene dynamics derived from quasi-linkage equilibrium \cite{neher2011statistical}, we obtain $\theta=\frac{2}{r}$, larger than $
\Theta$. This is because the sampling space of the inter-bacteria HGT tends to concentrate on the genotypes containing abundant genes. The biased sampling can be written as a quadratic term in the mean field approximation, resulting in stronger fluctuations at the gene level than the genotype level (see SI Sec. \ref{app:regimeIII} for more details).

However, our previous analysis does not work for Regime II as either the phage or the bacterium can go extinct, and the Lyapunov function $E$ is no longer well defined. 
Fig. \ref{fig:gamma_distribution}ab shows that most strains have small population sizes, represented by a large pile-up near zero, while a few booming strains dominate in the system, represented by an exponential tail on the right. We can focus on the booming strains, and the exponential tail in the Gamma distribution suggests they have a typical population size (peak size) $\sim \Theta$ \cite{pearce2020stabilization}.

We next estimate the typical peak size of the booming strains by connecting the "kill-the-winner" mechanism \cite{xue2017coevolution, doebeli2021boom, maslov2017population} with the establishment probability in population genetics \cite{desai2007beneficial, neher2010rate} as follows: the booming bacterial strain has an approximately constant fitness $s$ as long as its corresponding phage population is small. Hence, the bacterial strain (after establishing itself with the population size $\frac{1}{2s}$) follows deterministic exponential growth $e^{st}$ until the phage with antitoxin genes that allow infection of this bacterial strain emerges due to HGT and is itself established \cite{desai2007beneficial} (see Fig. \ref{figS:branching_process1}). Since the boom-bust cycles show that, most of the time, the phage does not impede the exponential growth of the susceptible bacterial strain until its establishment, we can use branching processes to evaluate the phage's establishment probability conditioned on the susceptible bacterial strain's instantaneous abundance \cite{neher2014predicting}. Then considering the constant rebirth rate $rN/K$ for one specific phage strain along with its probability of establishment, we can estimate the waiting time for the first phage to get established \cite{desai2007beneficial}. Upon establishment, given the large abundance of the bacterial strain at this time, the established phage proliferates rapidly, leading to the immediate decline of the susceptible bacterial strain population. Consequently, the bacterial strain population size at the time of the phage's establishment can be used to determine its typical peak size. Our analysis (see SI Sec. \ref{app:regimeII} and Fig. \ref{figS:branching_process2} for details) shows that the peak size follows an exponential distribution with exponential rate $-2r$, yielding $\Theta \approx \frac{1}{2r}$ . 
In this regime, as $r$ is small, different booming strains are weakly correlated and the gene abundance distribution is approximated by the sum of $L-1$ independent Gamma distribution for genotypes, which is still a Gamma distribution with the same exponential tail, yielding $\theta =\Theta= \frac{1}{2r}$.

In summary, our analysis of the dynamical behavior suggests the following scaling:
\begin{equation}\label{eq:theta_scalings_main}
\begin{cases}
\Theta=\theta = \frac{1}{2r} \quad &\text{for boom-bust cycles}\\  
\Theta=\frac{1}{2r}, \quad \theta=\frac{2}{r}  \quad &\text{for stochastic oscillations}
\end{cases}
\end{equation}
which match well with numerics in Fig. \ref{fig:critical}ab. Fig. \ref{fig:critical}b also shows the system can exhibit mixed dynamics when $\theta$ changes smoothly from $\frac{1}{2r}$ to $\frac{2}{r}$ in the transition from Regime II to III, consistent with our theory.

\noindent {\bf Critical values of $r$:} 
The shape of the PDF for genes and for genotypes provides a natural criteria designating different regimes: the power exponent in eq. (\ref{eq:gamma_main}) must be positive in order to avoid the PDF diverging at $n_G=0$ or $n_g=0$, which requires that
\begin{subequations}\label{eq:criteria_theta_main}
\begin{align}
\frac{N}{K\Theta}-1&>0, \\
\frac{2N}{L\theta}-1&>0.
\end{align}
\end{subequations}

Otherwise, the mode of the PDF is at zero, corresponding to an extinction of the genotype or gene. Hence the equalities $\tfrac{N}{K\Theta}=1$ and $\tfrac{2N}{L\theta}=1$ demarcate transitions between I/II and II/III respectively. 
Given the scaling of $\Theta$ and $\theta$ in eqs. (\ref{eq:theta_scalings_main}), the minimal $r$ to maintain gene and genotype diversity are
\begin{subequations}\label{eq:critical_rs_main}
\begin{align}
 r^g_c &\propto L/N, \\
 r^G_c &\propto K/N.
 \end{align}
\end{subequations}
While this scaling is robust, the argument does not determine the pre-factor (which is order 1); gene and genotype abundances fluctuate strongly near respective transitions, invalidating the mean-field theory assumptions that we have made in treating HGT (see SI Fig. \ref{figS:scalling_diff_gene_transfer}). Overall, as shown in Fig. \ref{fig:critical}cd,  eqs. (\ref{eq:critical_rs_main}) agree well with the results of numerical simulations. Alternatively, we can define transitions in terms of $L$ as a function of the total population size $N$ and the HGT rate $r$. Equations (\ref{eq:critical_rs_main}) imply that $L_{max}$ and $K_{max}$ at the transitions are proportional to $rN$ –a result confirmed by numerical simulations in
Fig. \ref{fig:phase}c.

In the preceding analysis, we set the inter-bacteria and bacteria-phage recombination rates to be equal, $r_B=r_V$ and derived the critical phage-bacteria HGT rate from the perspective of phage-induced growth inhibition.  We now consider the case $r_B\neq r_V$, highlighting the importance of inter-bacterial HGT as a survival mechanism for "escaping" predation.
In the limit of high $r_V$ when the system resides in Regime III, only negligible inter-bacteria HGT is required to maintain stable oscillatory behavior.
To locate the boundary between Regimes I and II, we estimate the minimum inter-bacteria HGT rate required for bacterial populations to survive phage predation by acquiring a suitable defense gene. A bacterial clone with population size $B(t)$ acquires genes at a rate $r_B B(t)$ and so that the expected number of established resistant offspring lineages generated during the "bloom" event is $sr_B \int dt B(t)$ which, to avoid extinction, should be $>1$. We denote the growth phase of the "booming" bacterial clone as $\tau_{boom}$ (the period from a bacterial strain’s establishment to its peak abundance), and we estimate the time integral using the average size of the booming clone, $N_B/n_{boom}$, where $n_{boom}$ is the average number of the booming bacterial strains in the system.  This leads to the inequality: $N_Br_Bs\tau_{boom}>n_{\text{boom}}$. 
In order to remain in Regime II where all genes are present, $n_{boom}> L$. We estimate $\tau_{boom}$ as the time required for a bacterial strain to grow from the established population size $1/s$ to the typical booming size ($N_B/L$) with the growth rate $s$. This estimation gives $s\tau_{boom}\approx \log\frac{sN_B}{L}$, which is of order 1 correction, leading to the simple relation (verified in Fig. \ref{fig:hostHGT}):
\begin{equation}\label{eq:HGT_ratio_inequality}
r_B > L/N_B.
\end{equation}
It demonstrates that when the phage infection rate is sufficient yet not unreasonably high  to suppress dominant bacterial growth, the 'escape' rate for bacteria to maintain genetic diversity is necessary and comparable to the value given by Eq. \ref{eq:critical_rs_main}a. 
Notably, as it is derived at the critical point of defense gene extinction, thus this relation is dependent of $r_B$ and system dynamics are governed primarily by the bacterial 'escape' rate rather than the phage 'predation' rate at the boundary. 

\noindent {\bf Generalization:} 
The transition criteria defined above are quite general. As an example, we consider the transition to the gene maintenance regime for the case with "tripleton" genotypes, i.e. each genotype carrying three genes so that $K\propto L^3$ in contrast to $K\propto L^2$ for the  "doubleton" case analyzed above. 
Fig. \ref{fig:critical} and Fig. \ref{figS:tripleton} show that the 
HGT rate required to maintain genetic diversity only grows linearly with the number of genes, in contrast to the faster polynomial (in $L$) growth of the genotype space. We can understand this from our criterion for gene persistence, which gives $rN$ as the estimate for the number of booming bacterial strains (each with a typical population size of order $1/r$). This number must exceed $L$ so that there are enough booming strains to carry all genes to escape from selective sweep, regardless of the "doubleton" or "tripleton" structure of the genotype. 


As there are multiple ways for phages and bacteria to acquire genes \cite{srikant2022evo}, we also consider the effect of inter-phage HGT as the dominant mechanism for generating "new" phages  (see SI Sec. \ref{sec:diff_recomb}). The mean-field theory description of HGT as a constant source of genotypes still applies, and Fig. \ref{figS:scalling_diff_gene_transfer}  shows that eqs. (\ref{eq:critical_rs_main}) remain valid.   Interphage transfer creates necessary escape genotypes by coinfection of a host by multiple phages each carrying one of the complementary antitoxins needed. The "biased sampling" effect of the bacterium-phage HGT does not apply  and we would expect that the infection rate required to maintain the gene pool in a high-dimensional genotype space must be substantially higher compared to the case of phage-bacteria transfer. This situation (which is unlikely when $rN>L$), is of course mitigated by any low rate of bacterium-phage HGT or new antitoxin emergence via de novo mutation (not considered in the present analysis). 

Our previous analysis focused on the antisymmetric and identical phage-bacteria pairs. However, in natural ecosystems,  the phage/bacteria population size ratio $\rho_n\sim10$, and the burst size $\beta=\rho_n\rho_s \sim 10-100$. Our approximate analysis (SI Sec. \ref{sec:generaliztion}) reveals that increasing the burst size, the phage population size, and the phage-bacteria HGT rate generally strengthens the “kill-the-winner” mechanism and reduces the critical HGT rates required, 
which is validated through numerical simulations in Figs. \ref{figS:r_ratios} \& \ref{figS:critical_nonsymmetric}. We also demonstrate that our results are robust to the extension to non-identical phage-bacteria pairs with parameters sampled from log-normal distributions  (see Fig. \ref{figS:heter}). 

Finally, our findings extend beyond the one-phage-one-bacteria infection scenario (details provided in SI Sec. \ref{sec:mutltiple_strain_infection}). 
We constructed multiple-strain infection networks by initializing the diagonal binary infection matrix, and then generalized the interaction to allow a single phage strain to simultaneously infect multiple bacterial strains, with the number of infections drawn from a Poisson distribution.  
Fig. \ref{figS:sparse_interaction} in SI demonstrates the robustness of our results over a wide range of sparse infection matrices. Thus we anticipate that our results can be applied to marine microecosystems, where co-isolation of phages and bacteria from the open ocean has revealed sparse infection profiles, with few broadly infective phage or resistant bacteria \cite{kauffman2022resolving}.

\vspace*{0.1cm}
\noindent {\bf Discussion:} 
In this article, we have formulated and explored a simple model of phage-bacteria competition driven by bacterial defense and phage counter-defense systems, focusing on the role of HGT in maintaining gene diversity. Our model aims to explain the stability of diverse immune repertoires in bacterial populations as they coevolve with bacteriophages. Although our model is easiest to directly relate to toxin-antitoxin systems, genome cataloging has identified that Restriction-Modification (RM) and CRISPR-Cas systems are among the most abundant classes of defense systems in genomes across bacterial phylogeny \cite{tesson2022systematic, costa2024accumulation}. Due to the diversity of proteins and domains that form TA systems, cataloging efforts do not properly account for the proportion of all the TA systems in phage defense, but an increasing number of TA system types continue to be implicated \cite{ernits2023structural, leroux2022toxin}. Like the presence of antitoxins on phage genomes to overcome immune barriers, modification enzymes homologous to those in RM systems  \cite{murphy2013bacteriophage} and CRISPR spacer-like elements \cite{camara2023bacteriophages} have been identified on phage genomes to overcome RM and CRISPR-Cas immunity. Further, it is expected that the source of these genes is from HGT-mediated acquisition from the RM and CRISPR systems of the bacterial host as part of the arms race. Thus the most abundant systems involved in immunity contribute to the arms-race evolution of phage-bacteria both in the distribution of the pan-immune system over the bacterial population and in the presence of counter-defense genes across phage genomes, consistent with our model of maintenance of genetic diversity through phage-bacteria gene transfer.



As described, our model is limited to understanding the “short-term” evolutionary process in phage-bacterial systems.  Our analysis focused on the feasibility of maintaining diversity in the presence of frequent extinction (of individual strains and their genotypes) solely by the virtue of HGT without the effect of mutation. 
The latter is of course well known to play an essential role in bacterium-pathogen coevolution, e.g., mutations in bacterial and phage proteins abrogate or restore their specific interaction \cite{rouzine2018antigenic, yan2019phylodynamic, marchi2021antigenic}. The evolution of TA systems is also thought to proceed by mutations, gene-swapping and possible de novo gene evolution, leading to rapid diversification and difficulty accounting for the number of systems associated with phage defense\cite{tesson2022systematic, ernits2023structural, leroux2022toxin}. Short, orphan genes in phage are associated with anti-immune function, and likely represent subsequent adaptation to overcoming bacterial defense \cite{ernits2023structural, otsuka2012dmd, srikant2022evo, tesson2024exploring}.


The importance of HGT processes in ecological dynamics and evolution of microbial communities is now broadly recognized \cite{arnold2022horizontal}.
HGT appears to be ubiquitous in nominally asexual microbes with an estimated rate comparable to the rate of mutation \cite{rosen2015fine}. Frequent HGT effectively unlinks genes from genomes, enabling selection to act on genes in the community metagenome, much like it does in linkage quasi-equilibrium in sexually reproducing populations \cite{neher2011statistical}. In this sense, the pool of TA genes forming the pan-immunity resource of the microbial community is maintained by selection through the boom/bust cycles of our Regime II. 

Although little is presently known of the rates of recombination and genome turnover in ecological settings, lab experiments of evolving phages have given us a window into the possibilities. In the lab, when maintaining a phage population size of $10^6-10^8$, it is possible to select evolved clones with recombination-based genome variation in 6-10 infection-generations \cite{srikant2022evo}. Such experiments have shown that phage genomes can amplify genes already present in their genome \cite{srikant2022evo}, acquire genetic material from co-infecting phage \cite{burrowes2019directed}, and from the infected bacterial cell’s genome \cite{Blower2013} to increase phage replicative fitness. Evolution experiments and screening for phage that escape from bacterial immune genes also readily identify point-mutations that occur at an appreciable rate in populations of similar size \cite{leroux2022toxin, stokar2023discovery, zhang2022direct}.

 To experimentally validate our claims, we can model extant genomes by placing TA systems on mobile genetic elements (MGEs) integrated into laboratory model bacteria, like E. coli. In a co-culturing experiment, by introducing naïve bacteria, we can measure the direct HGT transfer of these MGEs across bacteria. By adding infecting phage to the system, we can measure the effect on MGE spread across the bacterial population, the potential acquisition of antitoxin genes from TA locus by the phage genome to evolve resistance, and the ability to maintain diverse genomes during co-evolution.
 
While little known HGT mechanisms and processes constrain our model's predictive power, the "Red Queen" dynamics of such competitive coevolution can be folded into a suitably generalized model and would partially stabilize bacterial/pathogen strains by reducing the effective rate of extinction. This would effectively push the bacterial pathogen system towards the endemic coexistence in Regime III of our phase diagram.

Finally, we note that our model of pan-immunity is closely related to the multiple-islands-migration model in \cite{pearce2020stabilization}. Genes and genotypes are analogous to strains and "islands", respectively, and HGT plays a similar role as migration to stabilize the phage-bacteria dynamics. The spatial structure in their model is replaced with a genetic structure in ours, which allows for different applications. 

To conclude, our study provides a mathematical illustration of the pan-immunity hypotesis, showing how sufficiently high rate of HGT can ensure persistence of a diverse repertoire of defence genes distributed accross the pan-genome of bacteria and phages.  Combinations of diverse defense genes in the same genome result in a high-dimensional genotype space, leading to the emergence of numerous unoccupied "micro-niches", particularly when the interactions between defense and counter-defense genotypes are sparse. Our results imply that a reltively small number of thriving strains is sufficient to maintain diverse genes. These strains undergo rapid turnover among vast micro-niches, and the phage-bacteria arms race never ends.

\vspace*{0.1cm}
\noindent {\bf Acknowledgments:} The authors gratefully acknowledge stimulating discussions with Fridtjof Brauns, Tong Wang, Thierry Mora and Aditya Mahadevan.  WC acknowledges support via NSF PHY:1748958, GBMF Grant No. 2919.02, and Simons Foundation. BIS acknowledges support via NSF PHY:1707973 and NSF
PHY:2210612, S.S. is a Howard Hughes Medical Institute Awardee of the Life Sciences Research Foundation. WC also acknowledges generous help from Pankaj Mehta. The numeric results reported in this manuscript are performed on the Shared Computing Cluster of Boston University and UCSB Center for Scientific Computing. It is with deep sadness that we acknowledge the passing of Sriram Srikant, after this work was submitted for publication. Sriram’s passion for science, his breadth of knowledge, and his generosity of spirit remain an inspiration to us all.

\bibliography{ref.bib}

\clearpage
\renewcommand{\thefigure}{S\arabic{figure}}
\setcounter{figure}{0} 
\onecolumngrid
\begin{center}
\textbf{\large Supplementary Information}
\end{center}
\appendix
\renewcommand\appendixname{}
\renewcommand{\thesection}{\Roman{section}} 
\renewcommand\appendixname{}
\tableofcontents
\section{Summary of notation}\label{sec:summaryvariables}
We summarize the notation that appeared in the main text and supplemental information:

\vspace{3mm} 
\makebox[1.5cm]{$B$}   Bacteria genotype population size\par
\makebox[1.5cm]{$V$}   Bacterial phage (virus) genotype population size\par
\makebox[1.5cm]{$L$}  Number of genes. We always assume phages and bacteria have the same number of different genes\par
\makebox[1.8cm]{} $L$ ranges from $25$ to $50$ and $15$ to $25$ for "doubleton" and "tripleton", respectively.\par
\makebox[1.5cm]{$K$}  Number of genotypes, $K=\binom{L}{2} =L(L-1)/2$ for "doubleton"; for $K=\binom{L}{3}$  for "tripleton".  \par
\makebox[1.5cm]{$N_B$, $N_V$}  Total population sizes for the bacteria and phages. We set $N=N_B=N_V$ in the main text.  \par
\makebox[1.8cm]{} $N$ ranges from $10^5$ to $10^8$ in simulations.\par
\makebox[1.5cm]{$s$, $\omega$}   The bacterial intrinsic growth rate; the phage death rate. We set $s=\omega$ in the main text.  \par
\makebox[1.8cm]{} We always set $s=5\times 10^{-3}$.\par
\makebox[1.5cm]{$r_B$, $r_V$}    HGT rates for bacteria and phages. We set $r=r_B=r_V$ in the main text. \par
\makebox[1.8cm]{} $r$ ranges from $10^{-8}$ to $10^{-2}$ in simulations.\par
\makebox[1.5cm]{$n_G$}   Genotype abundance, no distinction for bacteria and phage\par
\makebox[1.5cm]{$n_G^*$}   Average population size for one specific genotype, $n_G^*=N/K$. If the bacterium and phage are assumed to have different fixed points, we use $n^*_B$ and $n^*_V$ to represent them. \par
\makebox[1.8cm]{} $n_G^*$ ranges from $10^2$ to $10^5$ in simulations. \par
\makebox[1.5cm]{$n_g$}   Gene abundance, no distinction for defense and counter-defense systems \par
\makebox[1.5cm]{$n_g^*$}   Average population size for one specific gene, $n_g^*=2N/L$ for "doubleton"\par
\makebox[1.5cm]{$\phi$}  The infection rate. We parameterize $\phi=\frac{s}{\rho_n n_G^*}$, which ranges from $10^{-8}$ to $10^{-4}$ in our simulations.  \par

\makebox[1.5cm]{$\beta$}  Burst size. $\beta=\rho_s \rho_n$, which usually ranges from 1 to 100 in our simulations. \par
\makebox[1.5cm]{$\Theta$}  Effective temperature for Boltzmann (Gamma) distribution of genotype abundances\par
\makebox[1.5cm]{$\theta$}   Effective temperature for Boltzmann (Gamma) distribution of gene abundances\par
\makebox[1.5cm]{$E$}   Lyapunov function, defined in eq. (\ref{eq:Lapu0})\par
\makebox[1.5cm]{$\rho_r$}  The ratio between the phage and bacteria HGT rates, $\rho_r = r_V/r_B$  \par
\makebox[1.5cm]{$\rho_s$}   The ratio between the phage's death rate and the bacterial intrinsic growth rate, $\rho_s = \omega/s$  \par
\makebox[1.5cm]{$\rho_n$}   The ratio between the total phage and bacterial population sizes, $\rho_n=N_V/N_B=n_V^*/n_B^*$ \par
\makebox[1.8cm]{} We typically range $\rho_r$, $\rho_s$ and $\rho_n$ from $0.1$ to $10$ in simulations.\par

\vspace{3mm} 
We would like to note that we assume bacteria and phage have anti-symmetric parameters so that they share the same statistical properties. If the bacterial and phage have different parameters, we use notations $X_B$ and $X_V$ ($X$ could be $s$, $\omega$, $N$, $r$).

\section{Simulation}\label{sec:simulation}
We must consider stochastic effects when the population size is finite. We use the $\tau$-leaping method \cite{gillespie2001approximate} to simulate eqs. (\ref{eq:sde_main}) in the main text:\begin{equation}\label{eq:sde_possion}
 \begin{aligned}
 B^{t+1}_{ij} = \text{Poisson}\left(B^{t}_{ij} e^{s(1-V^{t}_{ij}/n^*_V)}\right), \quad 
V^{t+1}_{ij} = \text{Poisson}\left(V^{t}_{ij} e^{ \omega(B^{t}_{ij}/n^*_B-1)}\right).  
\end{aligned}
\end{equation}
In the limit of strong HGT  and $s\sim \omega \ll 1$, we can write eqs. (\ref{eq:sde_possion}) into continuous equations,
\begin{equation}
  \begin{aligned}
\frac{dB_{ij}}{dt}&=s B_{ij}(1 - \frac{V_{ij}}{\rho_n n^*_G})+\sqrt{B_{ij}}\eta_{ij}^B,\\
\frac{dV_{ij}}{dt}&=\rho_s s V_{ij}( \frac{B_{ij}}{n^*_G}-1)+\sqrt{V_{ij}}\eta_{ij}^V.
\end{aligned}
\end{equation}
which is the LV dynamics of eqs. (\ref{eq:sde_main}) in the main text with demographic noise.
 Here $s$ and $\rho_s s$ are the host birth rate and the phage death rate.  $n_G^*$ and $\rho_n n^*_G$ are the host and phage characteristic population sizes which parameterize the infection rate and phage burst size; specifically $s/(\rho_nn^*_G)$ is the infection rate, and $\rho_s \rho_n$ is the phage's burst size. The last term is demographic noise, and $\eta^X$ ($X=B,V$) represents the unit white noise.

\begin{algorithm}[H]
\caption{Stochastic clone-based algorithm}\label{alg:agent}
\begin{algorithmic}[1]
	\State Set model parameters $L$, $s$, $\rho_s$, $r_B$, $r_V$,  $N_B$, and $N_V$.
	\State Initialize the population size $B_{ij}(t=0)$, $V_{ij}(t=0)$ for each strain (genotype) with $n^*_B$ and $n^*_V$, respectively. 
	\For{$t$ in $1:T$}
 \State {\bf Selection process}: 
      \State Each bacterial strain replicates itself following a Poisson distribution with the rates $B_{ij} e^{s[1-V^{t}_{ij}/(\rho_nn^*_G)] +\gamma(1-\frac{\bar{N}_B}{N_B})}$ , where $\bar{N_B} = \sum_{i<j} B_{ij} e^{s[1-V^{t}_{ij}/(\rho_nn^*_G)]}$. The term $\gamma(1-\frac{\bar{N}_B}{N_B})$ works as a regulator to constrain the total population approximately equal to $N$ \cite{neher2010rate}. We set $\gamma=0$ for no population constraint and $\gamma=\log{2}$ for the hard constraint.
    \State The phage replication follows a similar procedure. 
    \State {\bf Horizontal Gene transfer process}:
    \State Select arbitrary $r_BN_B$ individuals from all the existing bacteria and $r_VN_V$ individuals from all the existing phages, both via uniform sampling. 
    \State For the {\bf inter-bacteria} HGT case, then randomly select (via uniform sampling)  another $r_BN_B$ genes from the bacteria pool and replace one random gene from each of the selected $r_BN_B$ bacterial individuals with the selected genes. The {\bf inter-phage} and {\bf phage-bacteria} cases follow similar procedures. 
    \State The {\bf mean field approximation} case is equivalent to random mutations. To simulate this, we randomly mutate the selected genotype to an arbitrary genotype.
    \State Add the $r_BN_B$ new bacteria and $r_VN_V$ new phages back to the pool.
\EndFor
\end{algorithmic}
\end{algorithm}

All simulations are conducted with Julia. The codes are available on GitHub at \href{https:github.com/Wenping-Cui/GeneTransfer}{https:github.com/Wenping-Cui/GeneTransfer}.
\section{Canonical ensemble and Gamma distribution ansatz}\label{SI:canonical_ensemble}
In the deterministic limit, the antisymmetric Lotka-Volterra model for one bacterium-phage pair has the following conserved quantity, i.e., the Lyapunov function:
\begin{equation}\label{eq:Lapu0}
 E =(B -n^*_G\log{\frac{B}{n^*_G}}) +(V-n^*_G\log{\frac{V}{n^*_G}}),
\end{equation}
where the superscript $^*$ denotes the steady state. 

Our system consists of many bacteria-phage pairs coupled by gene exchange processes. Employing the idea of the canonical ensemble for interacting Lotka-Volterra systems \cite{kerner1957statistical},
the total Lyapunov function can be written as
\begin{equation}
E_{tot} = \sum_{i\neq j} (B_{ij} +V_{ij} -n^*_G \mathrm{log}\frac{B_{ij}}{n^*_G}-n^*_G \mathrm{log}\frac{V_{ij}}{n^*_G}).
\end{equation}
When all genes coexist for a long time, we may assume it is in \textit{thermal equilibrium} and $E_{tot}$ is conserved. It has been shown that the phase space has a unit of $\mathrm{log}B_i \mathrm{log}V_i$ to secure Liouville's theorem \cite{kerner1957statistical, goel1971volterra}; thus, we can derive the Boltzmann distribution from the principle of maximum entropy:
\begin{equation}
\begin{aligned}\label{eq:gamma_genotype0}
\rho(B, V)\dd \mathrm{log}B \, \dd \mathrm{log}V	&\propto  e^{-(B+V -n^*_G\log{\frac{B}{n^*_G}}-n^*_G\log{\frac{V}{n^*_G}})/\Theta} \dd \mathrm{log}B \, \dd \mathrm{log}V,
\end{aligned} 
\end{equation}
where $\Theta$ plays the role of an effective temperature. From the above expression, we can treat the bacterial and phage abundance distributions as independent and identical. We use $n_G$ to represent either the bacterial or phage abundances and obtain
\begin{equation}
\begin{aligned}\label{eq:gamma_genotype}
\rho(n_G) &\propto \left(\frac{n_G}{n^*_G}\right)^{n^*_G/\Theta}n_G^{-1} e^{-n_G/\Theta}.
\end{aligned} 
\end{equation}

After normalization, this is equivalent to the \textit{Gamma distribution}
\begin{equation}\label{eq:gamma}
 \rho(n_G) = \Gam{n^*_G}{\Theta}= \frac{n_G^{n^*_G/\Theta - 1} e^{-n_G/\Theta}}{\Theta ^{n^*_G/\Theta}\Gamma (\frac{n^*_G}{\Theta})},
\end{equation}
where  $\Theta$ is an unknown variable determined by system properties but can be evaluated numerically through
\begin{equation}\label{eq:theta}
\Theta = \left< (n_G - n^*_G)^2\right>/n^*_G,
\end{equation}
where the average $\left< \dots \right>$ is taken over the whole time series in the thermodynamic equilibrium phase with all strains surviving.

\subsection{Surviving fraction}
\begin{figure}[H]
\centering
\includegraphics[width=0.45\textwidth]{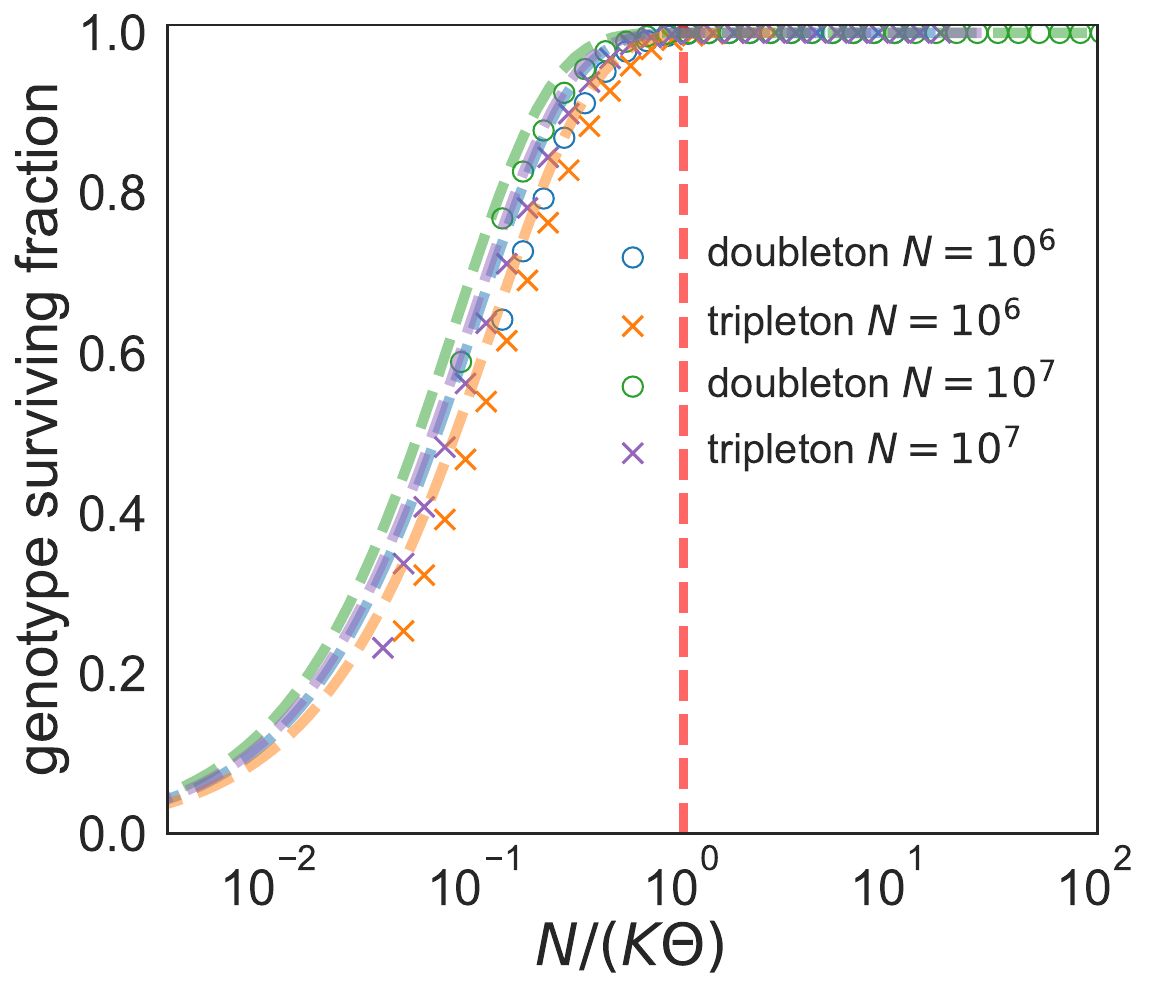}
\caption{Average surviving fraction of all genotypes over the whole time series. The vertical red dashed line is $N/(K\Theta)=1$. The remaining dashed lines represent theoretical predictions from eq. \ref{eq:extinction_ratio}, corresponding to the scatter points of the same colors. Here we obtain different $\Theta$ by adjusting the HGT rate $r$ with $L=40, s=0.005$. }
\label{figS:gamma_surviving}
\end{figure}

We can estimate the average surviving fraction of all possible genotypes over the time series by integrating the Gamma distribution from the natural cut-off 1 to infinity, yielding
\begin{equation}\label{eq:extinction_ratio}
f_{s} =\int_1^{+\infty}\frac{n_G^{n^*_G/\Theta - 1} e^{-n_G/\Theta}}{\Theta ^{n^*_G/\Theta}\Gamma (\frac{n^*_G}{\Theta})}d n_G= \Gamma \left(\tfrac{N}{K\Theta}, \tfrac{1}{\Theta}\right)/\Gamma (\tfrac{N}{K\Theta}).
\end{equation}
We would like to note that here $\Gamma \left(\tfrac{N}{K\Theta}, \tfrac{1}{\Theta}\right)$ is the incomplete gamma function to avoid confusion with the Gamma distribution notation in the main text. Fig. \ref{figS:gamma_surviving} shows eq. (\ref{eq:extinction_ratio}) matches remarkably well with numerical simulation. Additionally, it reveals that the fraction of surviving genotypes begins to rapidly decline from 1 when the rate of the power law component of the Gamma distribution falls below 1, signifying the onset of Regime II, where genotype extinctions occur.

\section{Stochastic LV model without horizontal gene transfer}\label{app:extinction} 
 
\begin{figure}
\centering
\includegraphics[width=0.85\textwidth]{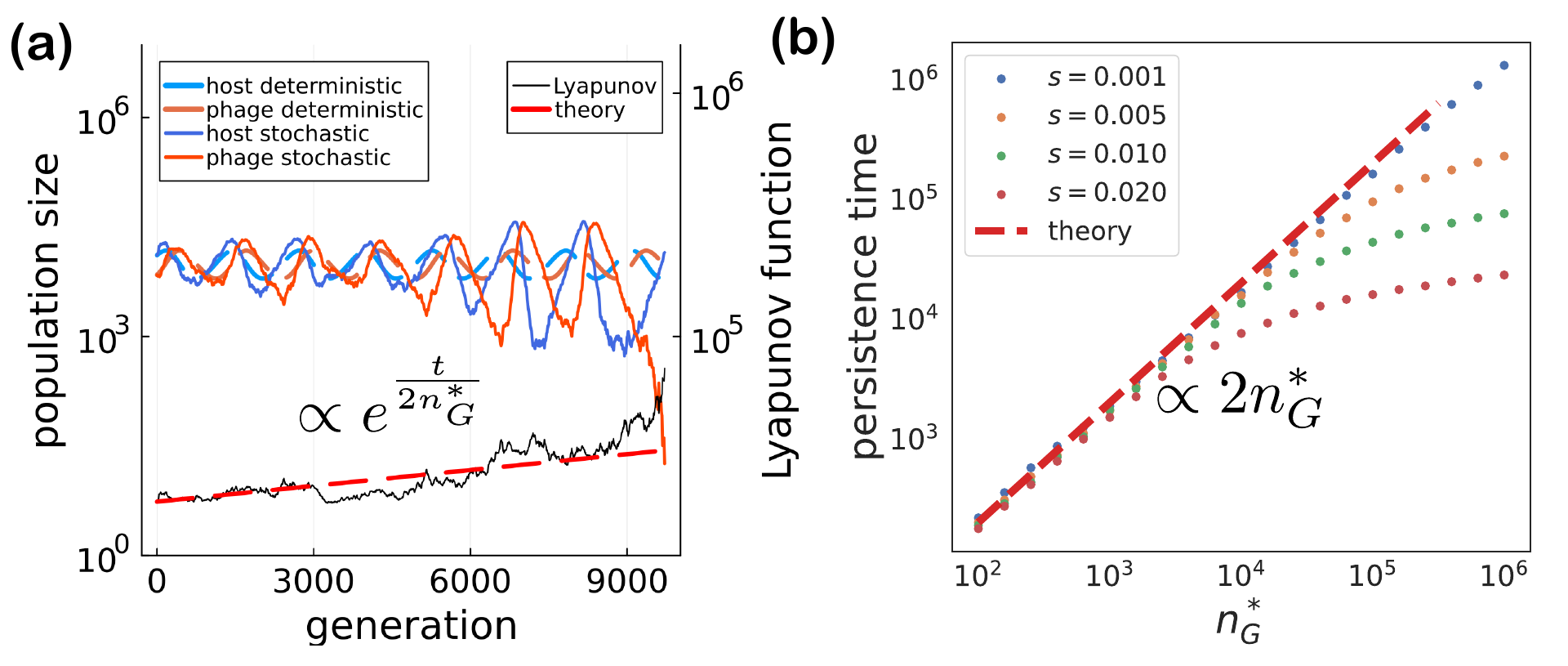}
\caption{{\bf (a)}  Left $y$ axis: deterministic and stochastic dynamics of one single pair of the bacteria-phage system. Right $y$ axis: the corresponding Lyapunov function and the theoretical prediction. The steady state for the phage and bacteria is $n^*_G=10^4$, and the intrinsic growth/death rate is $s=5\times 10^{-3}$; {\bf (b)} Persistence time for different $n^*_G$ and $s$. The data points are averaged over 1000 trials of simulation. }
\label{figS:extinction}
\end{figure}

In the deterministic limit, the bacteria and phage dynamics form a prey-predator oscillator that can oscillate forever. When the population size is finite, the stochasticity will drive the oscillation amplitude to increase until either the bacteria or the phage goes extinct, as shown in Fig. \ref{figS:extinction}a. In this section, we will derive the scaling of the extinction time. 

When the gene exchange rate is negligible compared the typical selection rate, 
\begin{equation}
 r \ll  s\left(1 -V/n^*_G \right)  \sim s \left(B/n^*_G  -1 \right)  \ll 1,
\end{equation}
we no longer consider HGT. Assuming the fitnesses for the bacterium and phage, $s\left(1 -V/n^*_G \right) $ and $s\left(B/n^*_G - 1\right)$, are much smaller than $1$, we can consider the stochastic LV equations
\begin{equation}\label{eq:StochasticLV}
\begin{aligned}
 \frac{dB}{dt} &=s B\left(1 -V/n^*_G \right) + \sqrt{B}\eta_B,\\
  \frac{dV}{dt} &=s V\left(B/n^*_G - 1\right)+ \sqrt{V}\eta_V,
 \end{aligned}
\end{equation}
where $\eta_B$ and $\eta_V$ are white noises with
$\left<\eta_B(t) \right> =\left<\eta_V(t) \right>=0$, $\left<\eta_B(t)\eta_V(t') \right> =0$, $\left<\eta_B(t)\eta_B(t') \right> =\delta(t-t')$, and  $\left<\eta_V(t)\eta_V(t') \right> =\delta(t-t')$. 

Instead of studying the two-dimensional nonlinear LV dynamics, we can study its stability through the Lyapunov function \cite{dobrinevski2012extinction} given by eq. (\ref{eq:Lapu0}).
Substituting eq. (\ref{eq:Lapu0}) into the deterministic part of eqs. (\ref{eq:StochasticLV}) shows that the Lyapunov function is conserved in the deterministic case, corresponding to a stable oscillator.

For the stochastic LV model, we can apply It\^{o}'s lemma \cite{gardiner1985handbook},
\begin{equation}
\begin{aligned}
dE  &=\left(\frac{\partial E}{\partial B}\frac{dB}{dt} +\frac{\partial E}{\partial V}\frac{dV}{dt} \right) dt+ \frac{1}{2}\left(\frac{\partial^2 E}{\partial B^2}B +\frac{\partial^2 E}{\partial V^2}V \right) dt + \sqrt{B}\frac{\partial E}{\partial B} d\eta_B+ \sqrt{V}\frac{\partial E}{\partial V} d\eta_V\\
 &=n^*_G\left( \frac{1}{2B} +\frac{1}{2V}\right) dt + \sqrt{\frac{(B-n^*_G)^2}{B} + \frac{(V-n^*_G)^2}{V}} d\eta\\
 &=\frac{n^*_G(E+n^*_G\log{\frac{B}{n^*_G}}+n^*_G\log{\frac{V}{n^*_G}})}{2BV}dt + \sqrt{\frac{(B-n^*_G)^2}{B} + \frac{(V-n^*_G)^2}{V}} d\eta
 \end{aligned}
\end{equation}
We can apply a linear-noise approximation around the steady state
to approximate it as
\begin{equation}
\begin{aligned}
dE  &\approx \left[\frac{E}{2n^*_G}-\frac{E(\Delta V + \Delta B)}{2(n^*_G)^2}+\frac{\Delta V + \Delta B}{2n_G^*} \right]dt + \sqrt{\frac{ \Delta B^2 + \Delta V^2}{n^*_G} }d\eta.
 \end{aligned}
\end{equation}
where  $B = n^*_G + \Delta B$ and  $V = n^*_G + \Delta V$.
In the limit $n^*_G\gg \Delta V, \Delta B$, we can keep the zeroth order, and the Lyapunov function grows exponentially with time:
\begin{equation}\label{eq:Lyapunov_weak}
E(t) \approx E(0)e^{\frac{t}{2n^*_G}}. 
\end{equation}
We do not know the threshold of the Lyapunov function when the oscillator becomes unstable.  However, from eq. (\ref{eq:Lyapunov_weak}),  we can roughly estimate the typical persistence time of the oscillator to be about $2n^*_G$. 

Fig. \ref{figS:extinction}a shows one simulation trial with the initial condition around the steady state, and the Lyapunov function grows approximately like eq. (\ref{eq:Lyapunov_weak}). Fig. \ref{figS:extinction}b compares the average persistence time with our theory, and $2n^*_G$ is a good estimation when the average clone size is below $10^5$. This is surprising because our result is based on the linear-noise approximation, which is no longer valid when the system is far from the steady state. Luckily, the amplitude of the oscillator increases dramatically after the linear-noise approximation fails, so our approximation is not bad.

\section{Regime III: stochastic oscillation with high HGT rates}\label{app:regimeIII}
\subsection{Stochastic Lotka-Volterra equations with HGT}
In this regime, the HGT rate is comparable to or larger than the fitness, i.e.,
\begin{equation}\label{eq:consistency_strong_recombination}
    s(-V/n_G^* + 1) \sim  s (B /n_G^*-  1) \lesssim r \ll 1,
\end{equation}
but both are still small compared with 1. 
We can write down the approximated Langevin equations:
\begin{equation}\label{eq:approx_SDE_mirgation}
\begin{aligned}
 \frac{dB}{dt} &= sB(-V/n_G^* + 1)+ rn_G^* + \sqrt{B}\eta_B,\\
  \frac{dV}{dt} &= s V(B /n_G^*-  1)+ rn_G^* + \sqrt{V}\eta_V.
 \end{aligned}
\end{equation}
The first term is the ordinary Lotka-Volterra dynamics. For the second term, we assume that the HGT process uniformly samples over the whole genotype space. This suggests HGT plays a similar role to migrations \cite{pearce2020stabilization, ottino2020population, agranov2021extinctions}. The third term comes from the demographic noise.

Applying It\^{o}'s lemma  \cite{gardiner1985handbook}, the stochastic differential equation of eq. (\ref{eq:Lapu0}) becomes
\begin{equation}
\begin{aligned}\label{eq:SDE_recombiantion}
dE  &=\left(\frac{\partial E}{\partial B}\frac{dB}{dt} +\frac{\partial E}{\partial V}\frac{dV}{dt} \right) dt+ \frac{1}{2}\left(\frac{\partial^2 E}{\partial B^2}B +\frac{\partial^2 E}{\partial V^2}V \right) dt + \sqrt{B}\frac{\partial E}{\partial B} d\eta_B+ \sqrt{V}\frac{\partial E}{\partial V} d\eta_V\\
 &= \left(\frac{n_G^*}{2B} +\frac{n_G^*}{2V} +r\frac{n_G^*(B-n_G^*)}{B} +r\frac{n_G^*(V-n_G^*)}{V} \right) dt + \sqrt{\frac{(B-n_G^*)^2}{B} + \frac{(V-n_G^*)^2}{V}} d\eta.
 \end{aligned}
\end{equation}

\subsection{Self-consistency relations}
\subsubsection{Solution of $\Theta$ and critical point}
In this regime, we hope that the time average of the Lyapunov function does not change. We then evaluate the drift part averaged with the canonical ensemble
\begin{equation}
 \int \rho(B)\rho(V)\left(\frac{n_G^*}{2B} +\frac{n_G^*}{2V} +r\frac{n_G^*(B-n_G^*)}{B} +r\frac{n_G^*(V-n_G^*)}{V} \right) dB dV= \frac{n_G^*}{(n_G^*-\Theta)} -\frac{2rn_G^*\Theta}{n_G^*-\Theta}=0.
\end{equation}
where $\rho(B)$, $\rho(V)$  are  Gamma distributions defined in eq. (\ref{eq:gamma_genotype}).
The self-consistency relation gives
\begin{equation}\label{eq:Thermodynamics_self_consistency}
\Theta = 
\begin{cases}
\frac{1}{2r}, \quad &2rn^*_G - 1>0, \\
\text{no solution}, \quad  &2rn^*_G - 1\leq 0.
\end{cases}
\end{equation}
Our analytical solution suggests there is a transition at $2rn^*_G =1$. The underlying reason is that the left tail of genotype Gamma distribution follows a power law with the exponent $2rn^*_G-1$. When the power exponent is below zero, $\rho(n_G)$ diverges at $n_G\rightarrow 0$, suggesting the phage and bacterium can go extinct, and only a fraction of genotypes are present in the system. Thus, the critical transition between Regime II and III happens at 
\begin{equation}
r^G_c = \frac{1}{2n^*_G}=\frac{K}{2N}.
\end{equation}
\subsubsection{An alternative explanation from LV dynamics}
We can see HGT processes help stabilize the system in the deterministic term in eq. (\ref{eq:SDE_recombiantion}). We want to ensure that the deterministic term is negative when the oscillation amplitude is large. Regarding the symmetry between $B$ and $V$, we only consider the case $B\ll n_G^*$ and $V\gg n_G^*$ and ignore the phage part in the term. Then we obtain the stability condition:  
$\frac{n_G^*}{2B} -r\frac{(n_G^*)^2}{B} <  0$, yielding
\begin{equation}\label{eq:thetabound}
2rn_G^* - 1>0,
\end{equation}
which provides an alternative view of the critical transition at $r_c^G$, suggesting the predictions of $r_c^G$ from canonical ensemble and stochastic LV dynamics are consistent with each other.

\section{Regime II: boom-bust cycles with low HGT rates}\label{app:regimeII}
When the HGT rate is weak, some phage or bacterial genotypes go extinct and take a long time to return to the ecosystem. In this regime, there are boom-bust cycles as in Fig. \ref{fig:gamma_distribution}a, and the continuous stochastic Lotka-Volterra equation is no longer valid. Since the boom-bust cycles show that, most of the time, the phage does not impede the exponential growth of the susceptible bacterial strain until its establishment,
we can apply a \textit{Branching Process} analysis to this regime.

The rising bacterium has an approximately constant fitness $s$ as its corresponding phage has a small population size. Fig. \ref{figS:branching_process1}a shows that after the rising bacterium gets established, it will grow exponentially with rate $s$ until the corresponding phage gets established.

\subsection{Establishment (fixation) probability in a changing environment}
\begin{figure}
\centering
\includegraphics[width=0.75\textwidth]{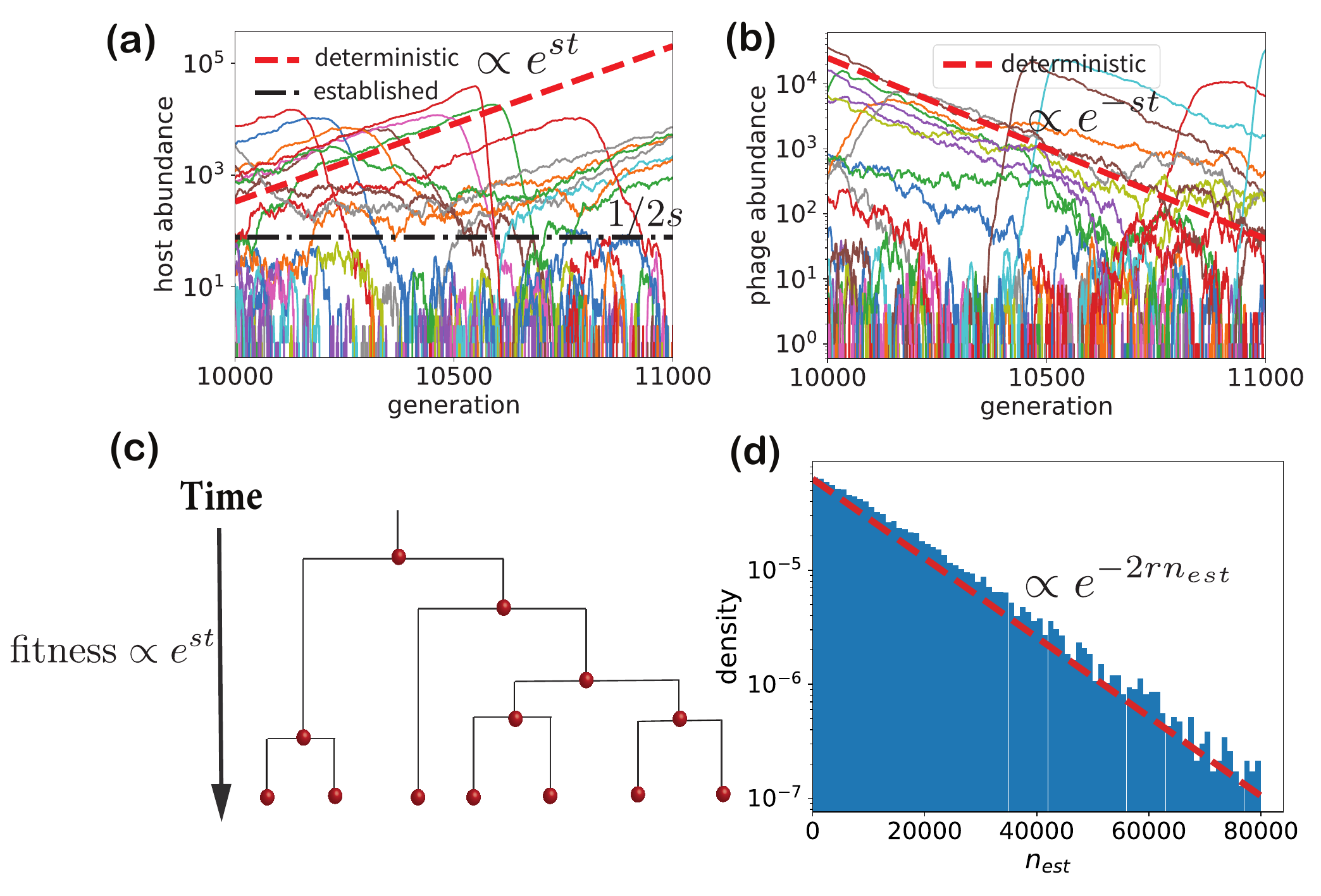}
\caption{{\bf (a)} A snapshot of multiple-bacteria population dynamics. The black dotted-dash line is the typical established population size $\frac{1}{2s}$ as a boundary between stochastic fluctuation and deterministic growth $\propto e^{st}$ (the red dashed line). {\bf (b)} A snapshot of multiple-phage population dynamics. The red dashed line shows the phage abundance almost decreased $\propto e^{-st}$ deterministically after dramatically booming. {\bf (c)} A schematic illustrating a branching process of one strain of phages with its fitness increasing exponentially in time as its susceptible bacterium follows exponential growth. {\bf (d)} The distribution of the bacterium abundance at the moment when the phage gets established and starts booming follows our theoretical prediction: an exponential distribution $\propto e^{-2rn_{est}}$.}
\label{figS:branching_process1}
\end{figure}
Assuming the bacterium starts to grow exponentially at $t=0$ with some established population size $n_0$, the phage's fitness $x$ follows
\begin{equation}\label{eq:phagefitness}
x(t) =  s\left( \frac{n_0}{n_G^*} e^{s t} -1\right).
\end{equation}
Eq. (\ref{eq:phagefitness}) shows the phage only has a positive fitness after the bacterium exceeds the mean population size,  and thus we are only interested in the time after the bacterium population size exceeds $n_G^*$. We set the time when the bacterium population size equals $n_G^*$ as zero. We want to answer the question: after a duration $\tau$, there are corresponding phages reborn from HGT; what is the probability that they get established?  This can be solved by a branching process in a changing environment \cite{neher2014predicting} (also see the scheme at Fig. \ref{figS:branching_process1}c).

We rewrite the phage's fitness into
\begin{equation}
x(\tau) =   s(e^{s \tau}-1), 
\end{equation}
where  $\tau=0$, as we defined, is when the bacterium population exceeds its mean abundance.

We assume the extinction probability after $t$ generations for a phage sampled at a given time $\tau$ follows a branching process with two offspring:
\begin{equation}
w(t|\tau) = [1-(x(\tau)+2)\Delta t]w(t-\Delta t|\tau +  \Delta t) +\Delta t+(1+x(\tau)) \Delta t w(t-\Delta t|\tau+\Delta t)^2.
\end{equation}
It is easier to use the variable $x$ instead of $\tau$, where $x$ is the fitness at time $\tau$,
\begin{equation}\label{eq:excintction_bacterium}
w(t| x) = [1-(x+2)\Delta t]w(t-\Delta t, x + s(x+ s)\Delta t) +\Delta t + (1+x)\Delta t w(t-\Delta t, x + s(x+ s)\Delta t)^2.
\end{equation}
In the continuum limit,
\begin{equation}
\frac{\partial  w(t| x) }{\partial t} - s(x+ s) \frac{\partial  w(t|x) }{\partial x} =1-(x+2)w(t, x) +(1+x)w(t| x)^2.
\end{equation}

We can rewrite the above equation into the survival probability $\phi(t,x) = 1- w(t|x)$,
\begin{equation}
\frac{\partial  \phi(t,x) }{\partial t} - s(x+s) \frac{\partial  \phi(t,x) }{\partial x} = x\phi(t, x) -(1+x)\phi(t, x)^2.
\end{equation}
We are interested in the establishment probability, which is equivalent to the survival probability in the long, but finite, time limit. In this limit, $\phi(t,x)$ is not sensitive to time and can be written as $\phi(x)$, 
\begin{equation}
- s(x+ s) \frac{\partial  \phi(x) }{\partial x} = x\phi(x) -(1+x)\phi(x)^2.
\end{equation}
In the limit of  $sx\ll1$ and $x \gg  s$,
\begin{equation}
\phi(x) \approx \frac{x}{1+x}\approx \frac{ se^{s\tau}}{1+ se^{s\tau}}.
\end{equation}

In deriving the survival probability, we assumed there were two offspring in the birth-death process. Our simulation uses a Poisson process to sample the number of offspring, (see Sec. \ref{AP:correctPoisson}), which contributes a factor of 2. After correction, it becomes
\begin{equation}\label{eq:Establishment}
\phi(\tau) \approx \frac{2 se^{s\tau}}{1+2 se^{s\tau}}.
\end{equation}

\subsection{Sampling from HGT}
We assume the phage sampled from HGT follows a Poisson process with a constant event rate $\frac{r}{K}$ per individual per generation per strain. In principle, the HGT process depends on the time-varying gene abundance. However, Fig. \ref{figS:gene_genotype_dynamics}  shows the gene dynamics, as the sum of $L-1$ genotype dynamics, have much smaller fluctuations and are also weakly correlated with the genotype dynamics, so we can take the average value and assume the event rate is a constant. The probability for sampling $n$ phages of a specific genotype is given as
\begin{equation}
P(n) = \frac{e^{-\frac{r N}{K}}}{n!}\left(\frac{ r  N}{K}\right)^n.
\end{equation}

We must consider that the establishment probability varies at different generations. The probability of $k$ established phages born at generation $\tau$ is given by
\begin{equation}
\begin{aligned}
 P(k , \tau) &= \sum^{\infty} _{n=k} \frac{e^{-\frac{rN}{K}}}{n!}\left(\frac{rN}{K}\right)^n \binom{n}{k}\phi(\tau)^k(1-\phi(\tau))^{n-k}\\
 & = \frac{e^{-\frac{rN\phi(t)}{K} }}{k!}\left(\frac{rN\phi(t)}{K}\right)^k\sum^{\infty} _{n=k}  \frac{e^{-\frac{rN}{K}+\frac{rN\phi(t)}{K}}}{(n-k)!}\left(\frac{rN}{K}-\frac{rN\phi(t)}{K}\right)^{n-k}\\
  & = \frac{e^{-\frac{rN\phi(t)}{K} }}{k!}\left(\frac{rN\phi(t)}{K}\right)^k,
 \end{aligned}
\end{equation}
which is an inhomogeneous Poisson distribution \cite{desai2007beneficial, snyder2012random}.

We are interested in when the first phage becomes established between $T$ and $T+\Delta T$.  In our simulation, we set $\Delta T = 1$, and the survival probability after exactly $T$ generations follows an exponential distribution, 
\begin{equation}
P(T) = \frac{rN}{K}\phi(T)e^{-\frac{rN}{K}\int_0^T \phi(\tau) \dd\tau}.
\end{equation}
We can evaluate its exponential part by
\begin{equation}
\frac{rN}{K} \int_0^T \phi(\tau) \dd \tau  = \frac{rN}{K} \int_0^T \frac{2se^{s\tau}}{1+2se^{s\tau}}\dd\tau  = \frac{rN}{Ks}\log\left(\frac{1+2se^{sT}}{1+2s}\right) \approx  2r n_G^* e^{sT}.
\end{equation}
Here we employ the fact that a typical establishment time $T$ obeys $se^{sT}\ll 1$ because the phage fitness is small.

We can write down its approximated probability distribution and consider the normalization
\begin{equation}
P(T) \approx  2sr n_G^* e^{2rn_G^* + sT-2r n_G^* e^{sT}}.
\end{equation}
Interestingly, this is the half-truncated \textit{Gumbel distribution}.

We are interested in the bacterium population size when the first phage gets established; it follows (see Sec. \ref{app:transformation})
\begin{equation}\label{eq:Qdistribution}
\begin{aligned}
 Q(n_G) = \frac{1}{sn_G}P(\frac{1}{s}\log{\frac{n_G}{n_G^*}})= 2re^{-2r (n_G-n_G^*)},  \quad  n_G^*\leq n_G    <\infty. 
 \end{aligned}
\end{equation}
We examine the above equation with numerical simulations, and it fits well as shown in Fig. \ref{figS:branching_process1}d.
\subsection{Consistency with canonical ensemble}
To be self-consistent with the exponential tail of the Gamma distribution in eq. (\ref{eq:gamma}), 
the effective temperature is given by
\begin{equation}
\Theta \approx \frac{1}{2r}.
\end{equation}
A question is raised as to why eq. (\ref{eq:Qdistribution}) follows the exponential distribution instead of the Gamma distribution. This is because we draw the histogram from the whole time series, which can be viewed as the sum of many independent and identical exponential distributions, naturally yielding the Gamma distribution. Our analysis matches well with simulation, as shown in Fig. \ref{fig:gamma_distribution} in the main text.

\subsection{Booming duration and time gap between consecutive booms}\label{app:boom_period}

\begin{figure}
\centering
\includegraphics[width=0.8\textwidth]{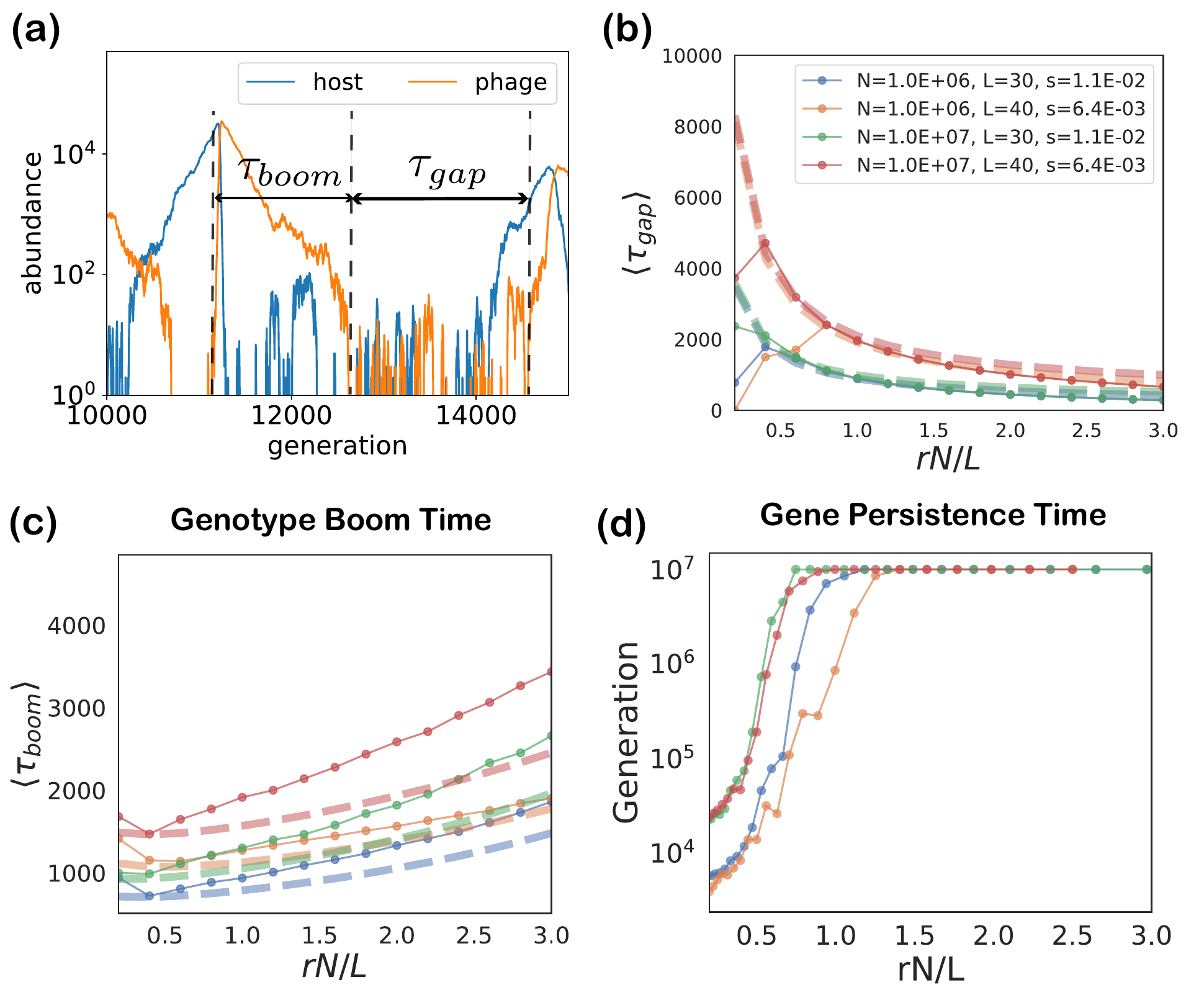}
  \caption{{\bf{(a)}} Scheme for the definition of the gap time $\tau_{gap}$ between two booms of phage and the phage booming time $\tau_{boom}$ in the boom-bust dynamics. 
  {\bf (b, c)} Comparison of the averaged $\tau_{gap}$ and $\tau_{boom}$ from simulation and our theoretical predictions (dashed lines). 
  {\bf (d)} Gene survival time for different $L$ and $N$ with a simulation time limit of $10^7$ generations.   }
\label{figS:branching_process2}
\end{figure}
It is interesting to investigate properties of the boom-bust cycles: the boom duration $\tau_{boom}$, defined as the time elapsed when the bacterium gets established ($n_{est}\sim \frac{1}{2s}$), reaches the peak ($n_G \sim\frac{1}{2r}$) and then dies out, $n_G=0$; and the gap time between two peaks, $\tau_{gap}$, shown in  Fig. \ref{figS:branching_process2}a. Because of time-reversal symmetry (perfect anti-symmetry), $\tau_{gap}$ and  $\tau_{boom}$ are identical for phages and bacteria.

Considering the bacterium, we can estimate the booming duration by the following equation:
\begin{equation}
\tau_{boom} = \frac{1}{s}\log \frac{s}{r} + \frac{4rn_G^*}{s(1- 2rn_G^*)}\log{\frac{1}{2r}}.
\end{equation}
The first term describes the exponential growth time from the established population size to the typical boom size. The second term describes two sequential processes: the phage reaches its boom size, and then, the bacterium drops to zero. We assume the two processes have the same typical time scale, contributing a factor of 2.  

$\tau_{gap}$, the time between when the bacterium/phage goes extinct and when it gets established, can be written as
\begin{equation}
\tau_{gap} = \frac{1}{2s}(\frac{1}{rn_G^*} + \log \frac{s}{r} ).
\end{equation}
The first term describes the typical time for a bacterial strain with a survival probability $2s$ and a HGT rate $rN/K$ to get established \cite{desai2007beneficial}. The second term results from the correction that the bacterial boom is always followed by the phage boom, and the bacterium cannot get established when the phage is in the boom phase.

In the simulation, $\tau_{boom}$ for the bacterium is underestimated. This is because the bacterium has a higher chance to undergo continuous booms after the phage dies, but the phage dies immediately after the bacterium dies. In Fig. \ref{figS:branching_process2}bc, we show our analytical estimation agrees well with the simulation.

Fig. \ref{figS:branching_process2}d gives a higher resolution of $r$ near the transition between Regime I and II. $rN/L = 1$  is the critical boundary given by eqs. (\ref{eq:criteria_theta_main}) in the main text.  It shows when passing $r_c^g$, i.e., $rN/L>1$, the strain boom time does not vary much, but the gene persistence time can increase dramatically,  much faster than exponential growth.

\subsection{Mean population size without the population size constraint}\label{app:population_size}
In Regime III, the prey-predator oscillator oscillates around the steady state $n^*_G$, causing the mean population size over the time series to also be $n^*_G$. However, in the boom-bust dynamics, for instance, the bacterium undergoes exponential growth and catastrophic recession, which does not have a steady state. We choose a hard constraint for the total population size and hence the average must be equal to $n^*_G$. 

The reader may be curious about the average clone size of each strain if we remove the hard constraint about the total population size. 
The gap period between two booms is given by
\begin{equation}
\tau_{gap} \approx \frac{1}{2s rn_G^*},
\end{equation}
which leads to 
\begin{equation}
    \left< n \right> = \frac{\int_{0}^{\frac{1}{s}\log\frac{s}{r}} dt \frac{1}{2s}e^{st}}{\frac{1}{s}\log\frac{s}{r} + \tau_{gap}} = \frac{\frac{1}{2sr}- \frac{1}{2s^2}}{\frac{1}{s}\log\frac{s}{r} + \frac{1}{2s rn_G^*} }\approx n^*_G.
\end{equation}
The above equation is a rough estimation. It needs careful treatment when the phages and bacteria are not perfectly anti-symmetric. 
\section{Different mechanisms of horizontal gene transfer}\label{sec:diff_recomb}
For the gene abundance distribution, it is helpful to normalize the population size to its fraction:
\begin{equation}\label{eq:approx_SDE}
\begin{aligned}
 \frac{dP_{ij}}{dt} &=sKP_{ij}\left(\frac{1}{K}- Q_{ij}\right) + J^P_{ij} +\sqrt{\frac{P_{ij}}{N}}\eta_P,\\
  \frac{dQ_{ij}}{dt} &=sKQ_{ij}\left( P_{ij}-\frac{1}{K}\right) +   J^Q_{ij}+ \sqrt{\frac{Q_{ij}}{N}}\eta_Q,
 \end{aligned}
\end{equation}
where  $P_{ij}=\frac{1}{N}B_{ij}$, $Q_{ij}=\frac{1}{N}V_{ij}$ are the fraction of bacterium/phage genotypes carrying genes $i, j$ so that the average of the off-diagonal elements $\left<P_{ij}\right>=\left<Q_{ij}\right>=1/K$. We set $P_{ii}=Q_{ii}=0$, $P_{ij}=P_{ji}$, and $Q_{ij}=Q_{ji}$ as each individual carries two different genes, and the order of genes does not affect the genotype. In other words. $P_{ij}$ and $Q_{ij}$ are symmetric matrices with diagonal elements equal to zero. $ J^X_{ij}$, represents the influx of clones for a specific genotype $X_{ij}$, and their forms depend on the detailed HGT processes. 

We would like to emphasize that the scaling of $\theta$ and $\Theta$ is also changed after normalization of the population size. As a result, we define $\bar{\theta}$ and $\bar{\Theta}$ for the normalized genotype and gene abundances, which obey
\begin{equation*}
   \theta = N  \bar{\theta}, \quad  \Theta = N  \bar{\Theta}.
\end{equation*}
\begin{figure}
\centering
\includegraphics[width=0.76\textwidth]{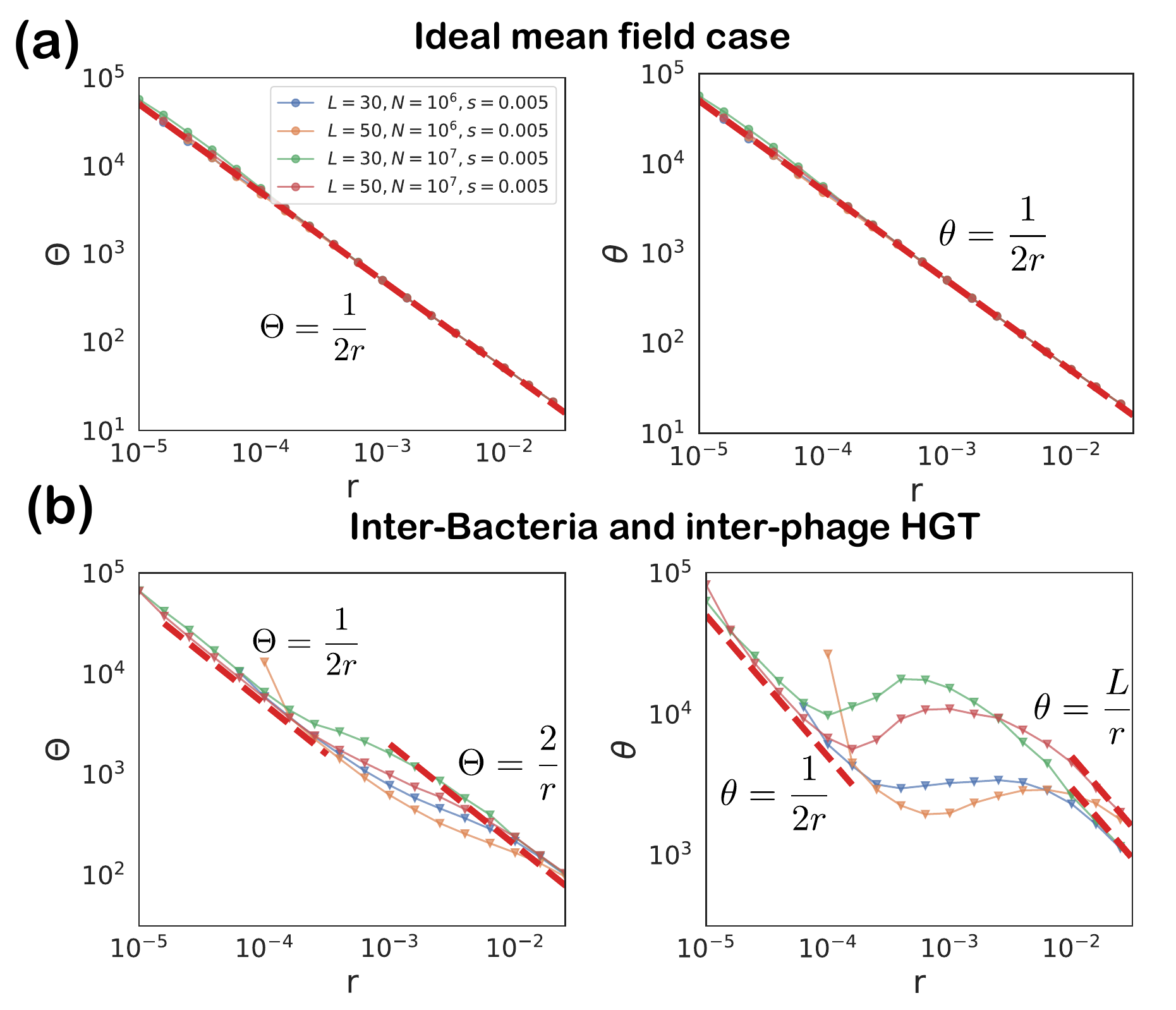}
\caption{Effective genotype and gene temperature for {\bf(a)} the ideal mean field case,  {\bf(b)} inter-bacteria and inter-phage HGT processes. The red dashed lines are theoretical predictions. }
\label{figS:diff_recombination}
\end{figure}

\subsection{Ideal mean-field case}
In this ideal case, we assume the rate of genotype generation is constant so that the clone influx is also constant. The dynamics are
\begin{equation}\label{eq:approx_SDE_mirgation_genotype}
\begin{aligned}
 \frac{dP_{ij}}{dt} &=sKP_{ij}\left(\frac{1}{K}- Q_{ij}\right) + \frac{r}{K}  +\sqrt{\frac{P_{ij}}{N}}\eta_P,\\
  \frac{dQ_{ij}}{dt} &=sKQ_{ij}\left( P_{ij}-\frac{1}{K}\right) +  \frac{r}{K} + \sqrt{\frac{Q_{ij}}{N}}\eta_Q.
 \end{aligned}
\end{equation}

We are interested in the marginal probability of genes:  $p_i = \sum_{j}P_{ij}$ and $q_i = \sum_{j}Q_{ij}$. We would like to note that  
\begin{equation}
\sum_i p_i = \sum_i q_i = 2,
\end{equation}
as each individual carries two genes, so it is not normalized. The influx for one specific gene is
\begin{equation}\label{eq:fluxrel}
\begin{aligned}
j^p_{i} = \sum_j J^P_{ij}=\frac{2r}{L}, &\quad j^q_{i} = \sum_j j^q_{ij}=\frac{2r}{L},
\end{aligned}
\end{equation}
which are also constant in this ideal case. 

The dynamics of $p_i$, $q_i$ become
\begin{equation}
\begin{aligned}
 \frac{d p _i}{dt} &\approx s p_i- sK\sum_{j} P_{ij}Q_{ij} +\frac{2r}{L} + \sqrt{\frac{p_i}{N}}\eta_p,\\
  \frac{dq_i}{dt} &\approx sK\sum_{j} P_{ij}Q_{ij}  -s q_i+ \frac{2r}{L} +  \sqrt{\frac{q_i}{N}}\eta_q.
 \end{aligned}
\end{equation}

We can evaluate the second term, $\sum_{j} P_{ij}Q_{ij}$, by assuming $P_{ij}$ and $Q_{ij}$ are weakly correlated:
\begin{equation}
\begin{aligned}
\sum_{j} P_{ij}Q_{ij} \approx \frac{1}{L}p_iq_i.
\end{aligned}
\end{equation}

Combining the above relations, we have
\begin{equation}
\begin{aligned}
 \frac{d p _i}{dt} &\approx \frac{Ls}{2} p_i \left(\frac{2}{L} -q_i\right) + \frac{2r}{L} + \sqrt{\frac{p_i}{N}}\eta_p,\\
  \frac{dq_i}{dt} &\approx  \frac{Ls}{2} q_i \left( p_i - \frac{2}{L}\right)  +\frac{2r}{L} +  \sqrt{\frac{q_i}{N}}\eta_q.
 \end{aligned}
\end{equation}

Similar to eq. (\ref{eq:SDE_recombiantion}), we can calculate the effective temperature of the Gamma distribution for $p_i$, $q_i$, and then rescale back to its population size. This gives
\begin{equation}
    \theta = \frac{1}{2r}.
\end{equation}

\subsection{Inter-bacteria and phage-bacteria HGT}
In this case, we take the phage-bacteria and inter-bacteria HGT into consideration, and the inflow HGT terms in eq. (\ref{eq:approx_SDE}) become
\begin{equation*}
\begin{aligned}
J^p_{ij}=&\frac{r}{4}\sum_{j'\neq j }P_{ij'}p_j+  \frac{r}{4}\sum_{i'\neq i}P_{ij'}p_i  + \frac{r}{2}P_{ij}(p_i+p_j)  \\
=& \frac{r}{4}(p_i -P_{ij})p_j +  \frac{r}{4}(p_j -P_{ij})p_i  + \frac{r}{2}P_{ij}(p_i+p_j)\\
=& \frac{r}{2}p_ip_j + \frac{r}{4}P_{ij}(p_i+p_j).
 \end{aligned}
\end{equation*}
Looking at the the first line of the above equation,  $rP_{ij'}p_j/2$ and $rP_{ij'}p_i/2$ describe the probability that one strain with one gene different, picks up  gene $i$ and transforms into genotype $ij$, noting that $p_i/2$ results from the normalization of $p_i$. It is then divided by another factor of 2 since we do not count the order of two genes.  $rP_{ij}(p_i/2+p_j/2)$ describes the probability that one genotype stays the same after the HGT. Similarly, we have
\begin{equation*}
\begin{aligned}
J^q_{ij}=\frac{r}{4}\sum_{j'}Q_{ij'}p_j+  \frac{r}{4}\sum_{i'}Q_{ij'}p_i  + \frac{r}{2}Q_{ij}(p_i+p_j) 
 = \frac{r}{4}q_ip_j + \frac{r}{4}q_jp_i + \frac{r}{4}Q_{ij}(p_i+p_j) .
 \end{aligned}
\end{equation*}

We can write the gene dynamics as
\begin{equation}
\begin{aligned}
 \frac{d p _i}{dt} &= \frac{Ls}{2} p_i \left(\frac{2}{L} -q_i\right) + \frac{r}{2}p_i\sum_{j} p_j +\frac{r}{4}p_i^2+\frac{r}{4}\sum_{j} P_{ij}p_j + \sqrt{\frac{p_i}{N}}\eta_p,\\
  \frac{dq_i}{dt} &=  \frac{Ls}{2} q_i \left( p_i - \frac{2}{L}\right)  +  \frac{r}{4}q_i\sum_{j} p_j +  \frac{r}{4}p_i\sum_{j} q_j  +\frac{r}{4}p_i q_i +\frac{r}{4}\sum_{j} Q_{ij}p_j +  \sqrt{\frac{q_i}{N}}\eta_q.
 \end{aligned}
\end{equation}

Utilizing $\sum_{j\neq i} p_j = 2 - p_i$ and $\sum_{j \neq i} q_j = 2 - q_i$, it can be reduced to 
\begin{equation}
\begin{aligned}
 \frac{d p _i}{dt} &= \frac{Ls}{2} p_i \left(\frac{2}{L} -q_i\right)  +\frac{r}{4}\sum_{j} P_{ij}p_j -\frac{r}{4}p_i^2 + rp_i + \sqrt{\frac{p_i}{N}}\eta_p,\\
  \frac{dq_i}{dt} &= \frac{Ls}{2} q_i \left( p_i - \frac{2}{L}\right)  +  \frac{r}{2}\left(p_i - q_i\right)+\frac{r}{4}\sum_{j} Q_{ij}p_j - \frac{r}{4}p_i q_i+ rq_i+  \sqrt{\frac{q_i}{N}}\eta_q.
 \end{aligned}
\end{equation}

\subsubsection{Decoupling of gene and genotype dynamics}
\begin{figure}[H]
\centering
\includegraphics[width=0.8\textwidth]{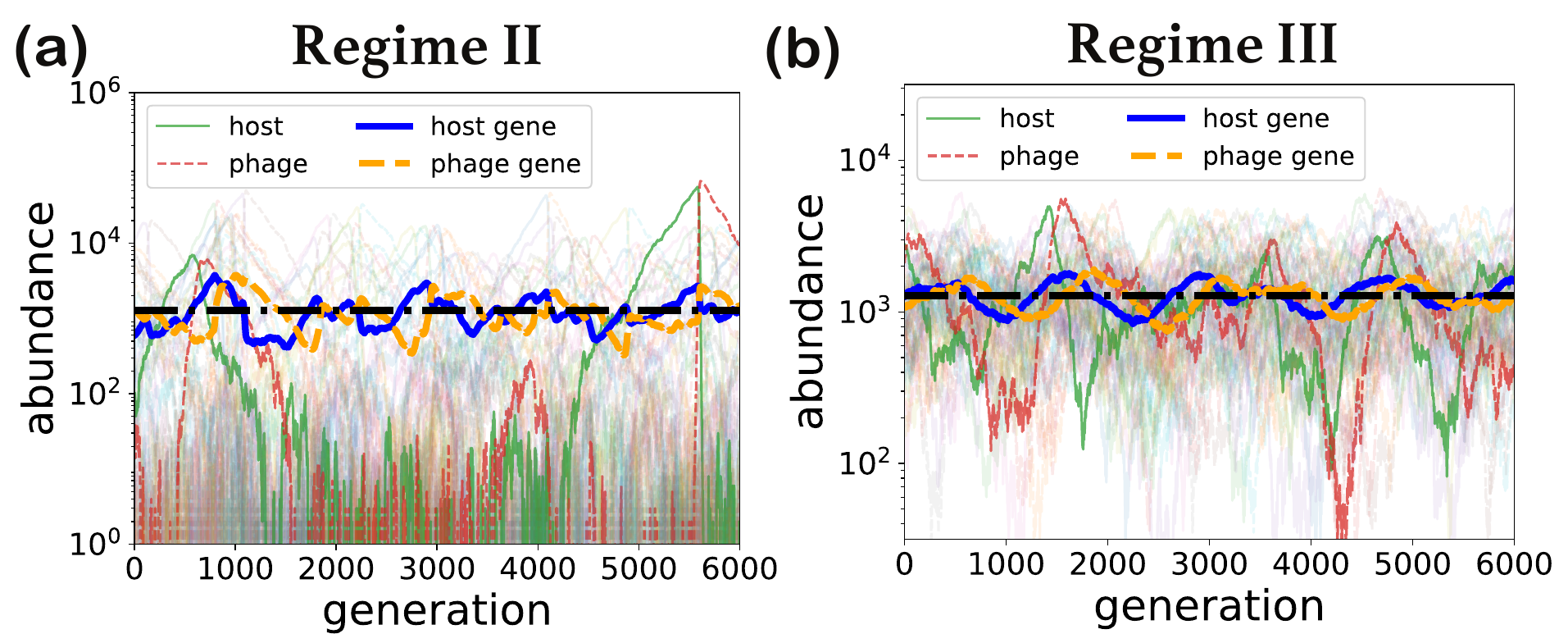}
\caption{ Comparison between gene and genotype  dynamics in Regime II {\bf(a)} and Regime III {\bf(b)}. We show the population dynamics of one specific gene and multiple genotypes containing that gene. In order to compare gene and genotype abundance at the same scale, the gene abundance is divided by $L-1$. The black dash-dotted line is the average of genotype abundances.  }
\label{figS:gene_genotype_dynamics}
\end{figure}

Fig. \ref{figS:gene_genotype_dynamics} suggests the different gene and genotype dynamics are weakly correlated. Hence we can average the gene dynamics out:
\begin{equation}
  \sum_{j} P_{ij}p_j \approx \frac{2}{L}p_i, \quad   \sum_{j} Q_{ij}p_j\approx \frac{2}{L} q_i.
\end{equation}

The mean-field gene dynamics become
\begin{equation}
\begin{aligned}
 \frac{d p _i}{dt} &= \frac{Ls}{2} p_i \left(\frac{2}{L} -q_i\right)  +\frac{r}{4}p_i(\frac{2}{L} -p_i) +r p_i+ \sqrt{\frac{p_i}{N}}\eta_p,\\
  \frac{dq_i}{dt} &=  \frac{Ls}{2} q_i \left( p_i - \frac{2}{L}\right)  + \frac{r}{2}\left(p_i - q_i\right)+\frac{r}{4}q_i(\frac{2}{L} -p_i)+r q_i+  \sqrt{\frac{q_i}{N}}\eta_q.
 \end{aligned}
\end{equation}

With $p^*=q^*=\frac{2}{L}$, similar to eq. (\ref{eq:SDE_recombiantion}), we can write down the dynamics of the Lyapunov function for the gene dynamics as:
\begin{equation}
\begin{aligned}
dE =& \left(\frac{p^*}{2Np_i} + \frac{p^*}{2Nq_i} -\frac{r}{4}(p_i-p^*)^2 -\frac{r}{2q_i}(q^*-q_i)^2 \right) dt \\
&+ (\frac{r}{2q_i}-\frac{r}{4})(p_i-p^*)(q_i-q^*) dt+ r(p_i-p^*) dt+ r(q_i-q^*) dt
+ \sqrt{\frac{(p_i-p^*)^2}{Np_i} + \frac{(q_i-p^*)^2}{Nq_i}} d\eta.
\end{aligned}
\end{equation}

The self-consistency relation yields
\begin{equation}
\begin{aligned}
&\int \rho(p_i)\rho_(q_i)\left[\frac{p^*}{2Np_i} + \frac{p^*}{2Nq_i} -\frac{r}{4}(p_i-p^*)^2 -\frac{r}{2q_i}(q^*-q_i)^2  \right] dq_i dp_i \\
 =& p^*\left[
\frac{1}{N(p^*-\bar{\theta})} -\frac{r}{4}\bar{\theta} -\frac{r\bar{\theta}}{2(p^*-\bar{\theta})} \right] =0.
\end{aligned}
\end{equation}
In the limit $rN\gg 1$, it gives
\begin{equation}
    \theta =N \bar{\theta}= N\left[\frac{L+1}{L} - \sqrt{\left(\frac{L+1}{L}\right)^2 -\frac{4}{rN}}\right]\approx \frac{2}{r}.
\end{equation}

\subsection{Inter-bacteria and inter-phage HGT}
Similar to the phage-bacteria HGT case, we can write down the mean-field gene dynamics as
\begin{equation}
\begin{aligned}
 \frac{d p _i}{dt} &= \frac{2s}{L} p_i \left(\frac{2}{L} -q_i\right)  +\frac{r}{4}p_i(\frac{2}{L} -p_i)+r p_i + \sqrt{\frac{p_i}{N}}\eta_p,\\
  \frac{dq_i}{dt} &= \frac{2s}{L}  q_i \left( p_i - \frac{2}{L}\right)   +\frac{r}{4}q_i(\frac{2}{L} -q_i)+r q_i+  \sqrt{\frac{q_i}{N}}\eta_q.
 \end{aligned}
\end{equation}

The dynamics of the Lyapunov function are characterized by 
\begin{equation}
\begin{aligned}
dE =& \left(\frac{p^*}{2Np_i} + \frac{q^*}{2Nq_i} -\frac{r}{4}(p_i-p^*)^2-\frac{r}{4}(q_i-q^*)^2  \right) dt\\
&+ r(p_i-p^*) dt+ r(q_i-q^*) dt+ \sqrt{\frac{(p_i-p^*)^2}{Np_i} + \frac{(q_i-p^*)^2}{Nq_i}} d\eta.
 \end{aligned}
\end{equation}
The self-consistency relation yields
\begin{equation}
 \int \rho(p_i)\rho_(q_i)\left[\frac{p^*}{2Np_i} + \frac{q^*}{2Nq_i} -\frac{r}{4}(p_i-p^*)^2-\frac{r}{4}(q_i-q^*)^2  \right] dq_i dp_i = \frac{p^*}{N(p^*-\bar{\theta})} -\frac{r}{2}p^*\bar{\theta} =0,
\end{equation}
which gives
\begin{equation}
\theta =N \bar{\theta}= N\left(\frac{1}{L} - \sqrt{\frac{1}{L^2} -\frac{2}{Nr}}\right)\approx \frac{L}{r}.
\end{equation}

\subsubsection{Effective genotype temperature}
The above result shows the effective gene temperature is proportional to $L$, much larger than the phage-bacteria case. This strong gene-gene correlation leads to a correction to $\Theta$, which is challenging to evaluate from the genotype dynamics as we can no longer decouple the gene and genotype dynamics.  

Thanks to the quasi-linkage equilibrium in the strong HGT regime \cite{neher2011statistical},  we can approximate the genotype fraction by the product of its marginal gene fraction, $P_{ij}\propto p_{i}p_{j},  Q_{ij}\propto q_{i}q_{j}$ and obtain that $P_{ij}, Q_{ij}$ follow a PDF:
\begin{equation}\label{eq:PDFprod}
\rho(x)=\frac{2x^{rN/L^2-1}K_0(2rN\sqrt{x}/L)}{(\frac{L}{rN})^{2rN/L^2}[\Gamma(rN/L^2)]^2},
\end{equation}
where $x$ represents either $P_{ij}$ or $Q_{ij}$, and $K_n$ is the modified Bessel function of the second kind. The above PDF has mean $1/L^2$ and variance $\frac{L^2+2Nr}{N^2L^2r^2}$. Then we can use eq. (\ref{eq:PDFprod}) to roughly estimate $\Theta$ for the Gamma distribution, which gives
\begin{equation}
 \Theta \approx \frac{L^2+2Nr}{Nr^2}\approx \frac{2}{r}.
\end{equation}
\begin{figure}
\centering
\includegraphics[width=0.8\textwidth]{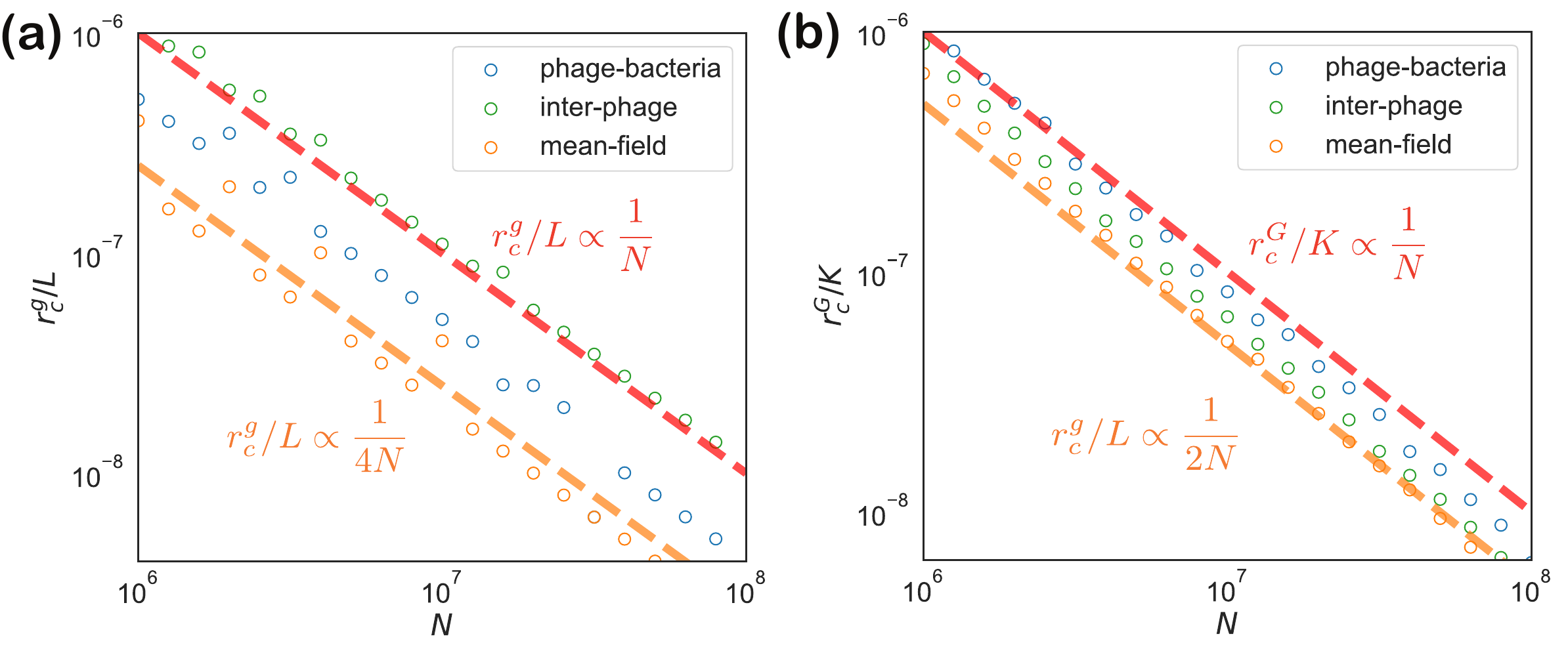}
\caption{Comparison of minimal HGT rates for (a) gene coexistence  and (b) genotype coexistence for different types of HGT: inter-bacteria and phage-bacteria HGT, inter-bacteria and inter-phage HGT. We also conduct simulations with constant rates of genotype generation from the mean-field approximation for comparison. As expected, our mean field theoretical prediction (the orange lines) matches perfectly with the simulation with constant rates of genotype generation from mean field estimation, but falls off with a factor within order 1 for the two different HGT processes (the red lines are theoretical predictions dropping the prefactor). }
\label{figS:scalling_diff_gene_transfer}
\end{figure}
\subsection{Summary}
We summarize the analytical results in Table \ref{table:summary_diff_gene_transfer} and show the comparison with numerical simulations in Fig. \ref{fig:critical}ab in the main text and Fig. \ref{figS:diff_recombination}. The scaling of $\Theta$ and $\theta$ shows different implementations of the HGT processes do not affect our criteria eqs. (\ref{eq:critical_rs_main}) much  (see Fig. \ref{figS:scalling_diff_gene_transfer}). We get the critical $r$ by comparing the extinction time of the first gene by running a simulation with a large threshold $T=250000$. We would like to note that for the ideal mean-field case, the extinct gene can always come back in the future, so the scaling of $r_c^g$ appearing in Fig. \ref{figS:scalling_diff_gene_transfer} works as an ideal case to compare with other cases and examine our theory. 

\begin{table}[H]
\begin{center}
\begin{tabular}{ |K{4cm}|| K{2cm} | K{2cm}| K{2cm} | K{2cm} | }
 \hline
    HGT &
\multicolumn{2}{c|}{Boom-bust cycles} &
\multicolumn{2}{c|}{Stochastic oscillator}  \\
& $\Theta$ & $\theta$ & $\Theta$ & $\theta$\\[0.1cm]
 \hline
mean-field & $1/2r$ & $1/2r$ &  $1/2r$    & $1/2r$\\ [0.1cm]
\hline
inter-bacteria/ phage-bacteria  & $1/2r$  & $1/2r$  &$1/2r$   & $2/r$\\[0.1cm]
\hline
inter-bacteria/ inter-phage  & $1/2r$  & $1/2r$ & $2/r$  & $L/r$\\[0.1cm]
 \hline
\end{tabular}
\end{center}
\caption{Summary of effective temperature for different types of HGT.\label{table:summary_diff_gene_transfer}}
\end{table}

\section{General parametrization}\label{sec:generaliztion}
\subsection{Distinct parameters for bacteria and phages}\label{sec:distinct_param}

In the main text, we set the ratios: $\rho_r=r_V/r_B$, $\rho_s=\omega/s$, and $\rho_n=n_V^*/n_B^*$ all equal to 1 to reduce the parameter space. In nature we often anticipate the phage to have a larger clone size $\rho_n>1$, a higher HGT rate $\rho_r>1$,  and for the burst size, proportional to $\rho_s$, to be also larger than 1. Let's investigate how they affect the effective temperature in Regime III first. 
\subsubsection{Regime III}
In this regime, the dynamics become
\begin{equation}
\begin{aligned}
 \frac{dB}{dt} &= s B\left(1 - \frac{V}{\rho_n n^*_G} \right) + rn^*_G  + \sqrt{B}\eta_B,\\
  \frac{dV}{dt} &= \rho_s s V(B/n^*_G - 1)+ \rho_r rb n^*_G + \sqrt{V}\eta_V,
 \end{aligned}
\end{equation}
where $r=r_B$, and $N=N_B$.

The Lyapunov function becomes
\begin{equation}
E =\rho_n(B -n^*_G\log{\frac{B}{n^*_G}}) + \frac{1}{\rho_s}(V-\rho_n n^*_G\log{\frac{V}{\rho_n n^*_G}}).
\end{equation}

We define the effective temperature $T$, and the bacterium, phage abundances follow 
\begin{equation}
 B\sim\Gam{n^*_G}{\Theta_B= T/\rho_n }, \quad V\sim \Gam{-\rho_n n^*_G}{\Theta_V=  \rho_s T}.
\end{equation}

Repeating the above steps, the self-consistency equation is
\begin{equation}
\left <\frac{ \rho_nn^*_G}{ 2B} +\frac{\rho_n n^*_G}{2\rho_s V} +\rho_nr\frac{n^*_G(B-n^*_G)}{B} + \rho_r r\frac{\rho_n n^*_G(V-\rho_n n^*_G)}{ \rho_s V} \right>_{\rho(B), \rho(V)}=0.
\end{equation}
After taking the average, it becomes
\begin{equation}
\begin{aligned}
\frac{\rho_s (\rho_n-2r  T)}{\rho_nn^*_G-T }+\frac{(1-2\rho_r \rho_s r  T)}{ \rho_nn^*_G -\rho_s T }&= 0.
\end{aligned}
\end{equation}
We drop the nonphysical solution, and the other solution of $T$ is a complicated expression:
\begin{equation}
T=\frac{\rho_n(\rho_s ^2+2 (1+\rho_r ) \rho_s   n^*_Gr+1)-\sqrt{\rho_n^2\left(\rho_s ^2+2 (\rho_r +1)  \rho_s  n^*_Gr+1\right)^2-8 \rho_n \rho_s  (\rho_s \rho_n +1) n^*_Gr (\rho_r  +\rho_s )}}{4 \rho_s  r (\rho_r  +\rho_s )}.
\end{equation}

\begin{figure}[H]
\centering
\includegraphics[width=0.95\textwidth]{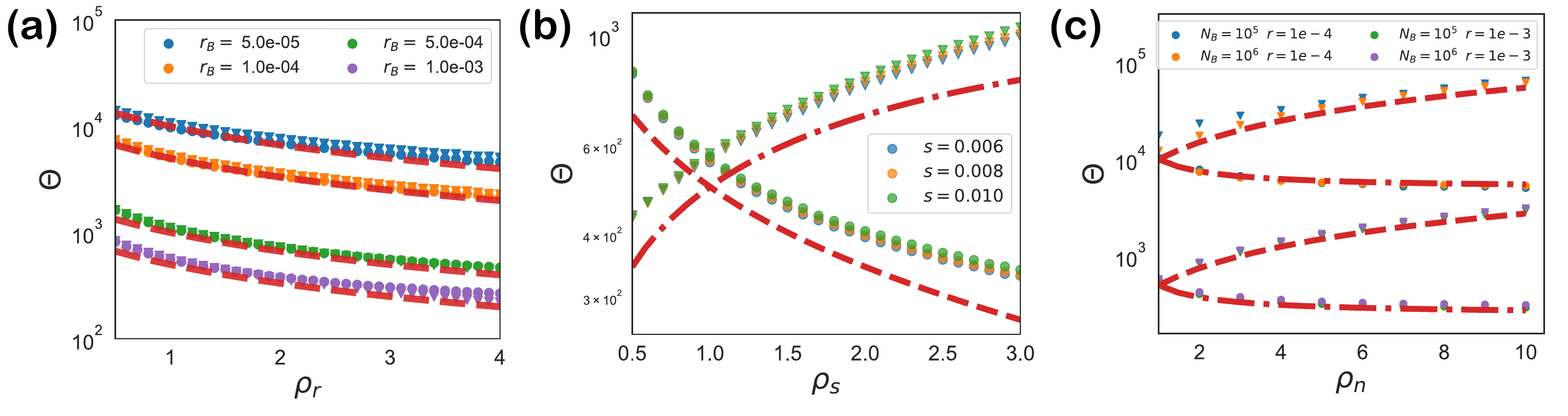}
  \caption{Effective genotype temperature $\Theta$ for bacteria and phages with {\bf(a)} different HGT rates, {\bf(b)} interaction strength and {\bf(c)} population size. The dashed lines and dash-dotted lines are predictions from the theory for bacteria and phages respectively. The scatter points are simulation data with $L=40$, $r=10^{-3}$, $s=5\times10^{-3}$, and $N=10^6$ unless specified. Circular and triangular markers represent bacteria and phages, respectively. }
\label{figS:diff_NrJ}
\end{figure}

If we only change one pair of parameters and keep the other two identical, we find simpler expressions:
\begin{itemize}
    \item Different gene exchange rate $r$: 
    \begin{equation}\label{eq:Theta_diff_r}
        \Theta_B = \Theta_V = \frac{1}{(1+\rho_r)r}.
    \end{equation}
    \item Different $N$: 
        \begin{equation}
      \Theta_B = \frac{T}{\rho_n}=\frac{1+\rho_n}{4b r}, \quad \Theta_V = T=\frac{1+\rho_n}{4r}.
        \end{equation}
    \item Different $s$ :
    \begin{equation}
        \begin{aligned}
                \Theta_B &=    \frac{1+4 \rho_s  n^*_G r+\rho_s ^2-\sqrt{1+2 \rho_s ^2+16 \rho_s ^2n^*_G r(n^*_G r -1)+\rho_s ^4}}{4 \rho_s  (\rho_s +1) r},\\  
  \Theta_V &=    \frac{1+4 \rho_s  n^*_G r+\rho_s ^2-\sqrt{1+2 \rho_s ^2+16 \rho_s ^2n^*_G r(n^*_G r -1)+\rho_s ^4}}{4  (\rho_s +1) r}.
        \end{aligned}
    \end{equation}
\end{itemize}

We examine the above results with simulation in Fig. \ref{figS:diff_NrJ}.

\subsubsection{Regime II}
We rewrite the phage's fitness, eq. \ref{eq:phagefitness},  into
\begin{equation}
x(\tau) =   \rho_s s(e^{s \tau}-1).
\end{equation}
The survival probability is given by
\begin{equation}
\phi(x) \approx \frac{x}{1+x}\approx \frac{ \rho_s se^{s\tau}}{1+ \rho_s se^{s\tau}}.
\end{equation}

The first passage time for the phage to get established is given by
\begin{equation}
P(T) \approx  2s \rho_r\rho_s \rho_n r n_G^* e^{2\rho_r\rho_s \rho_n rn_G^* + sT-2\rho_r\rho_s \rho_n r n_G^* e^{sT}}.
\end{equation}

We are interested in the bacterium population size when the first phage gets established; 
\begin{equation}
\begin{aligned}
 Q(B) = \frac{1}{sB}P(\frac{1}{s}\log{\frac{B}{n_G^*}})= 2\rho_r\rho_s \rho_n re^{-2\rho_r\rho_s \rho_n r (n_G-n_G^*)},  \quad  n_G^*\leq n_G    <\infty. 
 \end{aligned}
\end{equation}
which gives the temperature for the bacteria would be
\begin{equation}
    \theta_B = \frac{1}{2\rho_r \rho_s \rho_n r},
\end{equation}
It is not surprise that factors such as the phage-bacteria HGT rate, phage population size, and burst size promote the "kill-the-winner" mechanism. 

Then the critical "temperature" between Regime I/II is given by 
\begin{equation}
     N_B/\theta_B =2r_V N_B \rho_s \rho_n  \sim L. 
\end{equation}
However, this is not consistent with the collapse of the curves shown in Fig. \ref{fig:hostHGT}. This perhaps results from two reasons: 1. the interference among booming strains in the boom-bust dynamics introduces complex self-consistency relations between total population size and other model parameters. 2. the extreme population fluctuations in boom-bust cycles also leads to inaccuracy in stochastic simulations. We would leave them as the future work. Nevertheless, our analytical approach provides qualitative insights into the system's behavior across a broader parameter space.

\subsubsection{Summary}
\begin{figure}[H]
\centering
\includegraphics[width=0.95\textwidth]{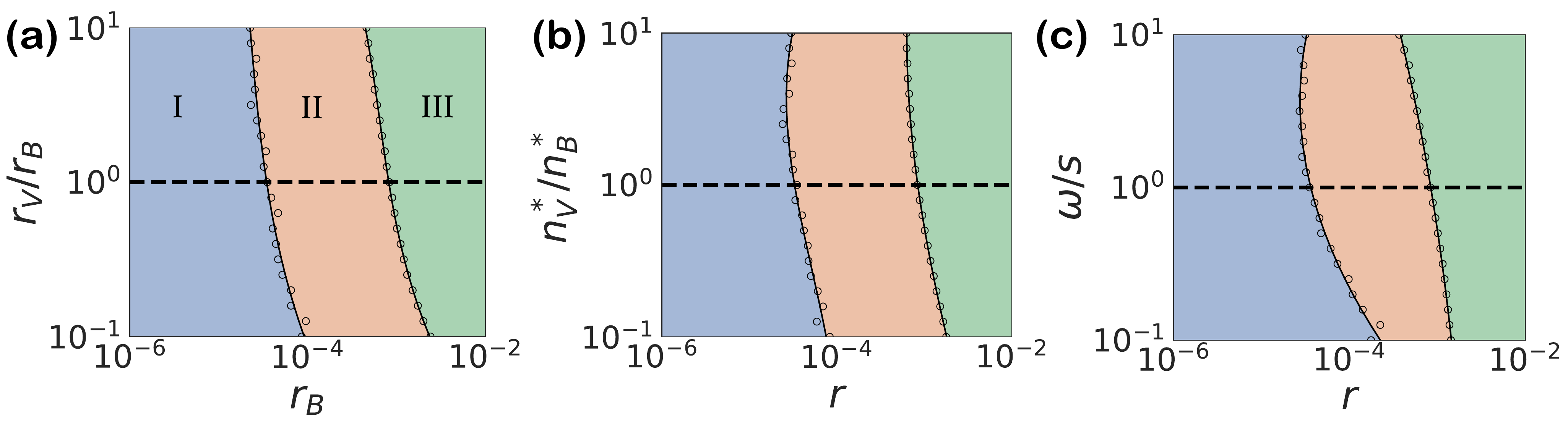}
\caption{Phase diagrams for phages and bacteria with different ratios of HGT rates, fixed points, and phage death rate/bacterial birth rate. The scatter points are the minimal HGT rates to keep gene and genotype diversity invariant over a simulation period $T=250000$. The black solid lines are spline interpolation through the scatter points. The horizontal dashed line corresponds to the simplified case shown in Fig. \ref{fig:phase} in the main text.
}
\label{figS:r_ratios}
\end{figure}

From the above calculations in Regime II\&III, we can see that these ratios cannot be simply canceled out through parameter rescaling because they may affect demographic noises and HGT processes non-trivially. However, qualitatively, when $\rho_r$, $\rho_s$, and $\rho_n$ are perhaps greater than 1, it implies the phage tends to kill the booming bacterium earlier, thereby reducing fluctuations in bacterial abundances and lowering the temperature. For instance, an increase in the phage's HGT rate or population size can enhance its establishment probability due to the larger rebirth rate. As a consequence, we can expect the setting of $\rho_r,\rho_s,\rho_n>1$ to shift the transition to a lower value of the bacterial HGT rate $r$, compared to the baseline case where $\rho_r$, $\rho_s$, and $\rho_n$ are set to 1, as shown in Fig. \ref{figS:r_ratios}. In essence, the natural settings ($\rho_r,\rho_s,\rho_n>1$) promote the gene and genotype diversity, compared to our simplified settings ($\rho_r=\rho_s=\rho_n=1$).  

\begin{figure}[H]
\centering
\includegraphics[width=0.7\textwidth]{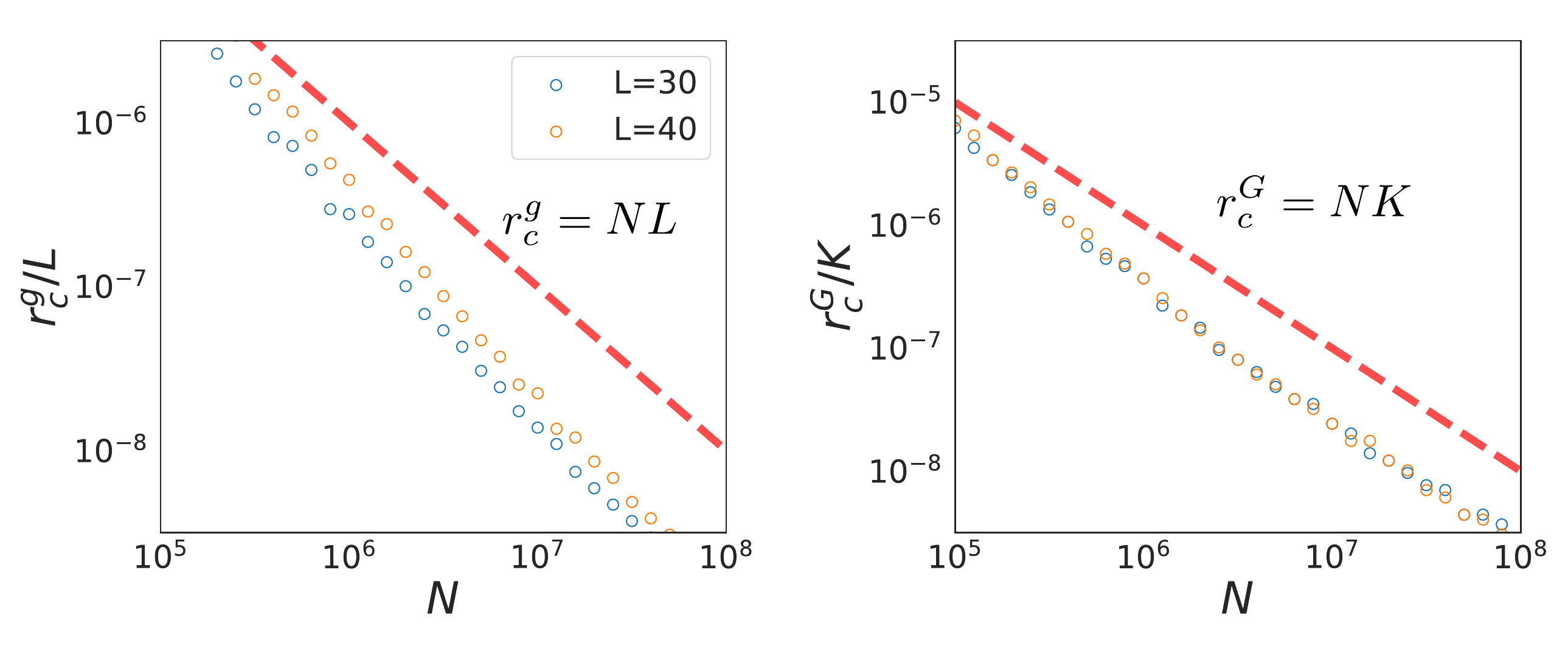}
\caption{Critical (minimal) HGT rates for gene and genotype coexistence for the non-symmetric parametrization. The simulation parameter set: the phage/bacteria population size ratio: $\rho_n=10$, the burst size: $\beta =\rho_s*\rho_n$ =100 and the phage/bacteria HGT ratio $\rho_r=0.1$.  }
\label{figS:critical_nonsymmetric}
\end{figure}

In nature, the phage/host population ratio can be as much as 10. For biological plausibility, we further test our theory with the phage/bacteria population size ratio: $\rho_n=10$, the burst size: $\rho_s*\rho_n$ =100 and the phage/bacteria HGT ratio $\rho_r=0.1$ (we have to carefully choose alpha otherwise the “killing-the-winner” mechanism is so strong that the bacteria is hard to extinct).  Fig. \ref{figS:critical_nonsymmetric} shows that our theory actually gives an upper bound of the critical HGT rates which the model with realistic parameters would fall below.

\subsection{Heterogeneous parameters}\label{sec:heterogeneous}
In the main text, we also assume that $n_G$ and $s$  are identical for all phage-bacteria pairs. We consider the heterogeneous case where $s$ and $n_G$ are sampled from the lognormal distribution for each phage-bacteria pair:
\begin{equation}
    \rho(x) = \frac{1}{\sigma x\sqrt{2\pi}}e^{-\frac{(\log x)^2}{2\sigma^2}}.
\end{equation}

\begin{figure}[H]
\centering
\includegraphics[width=0.85\textwidth]{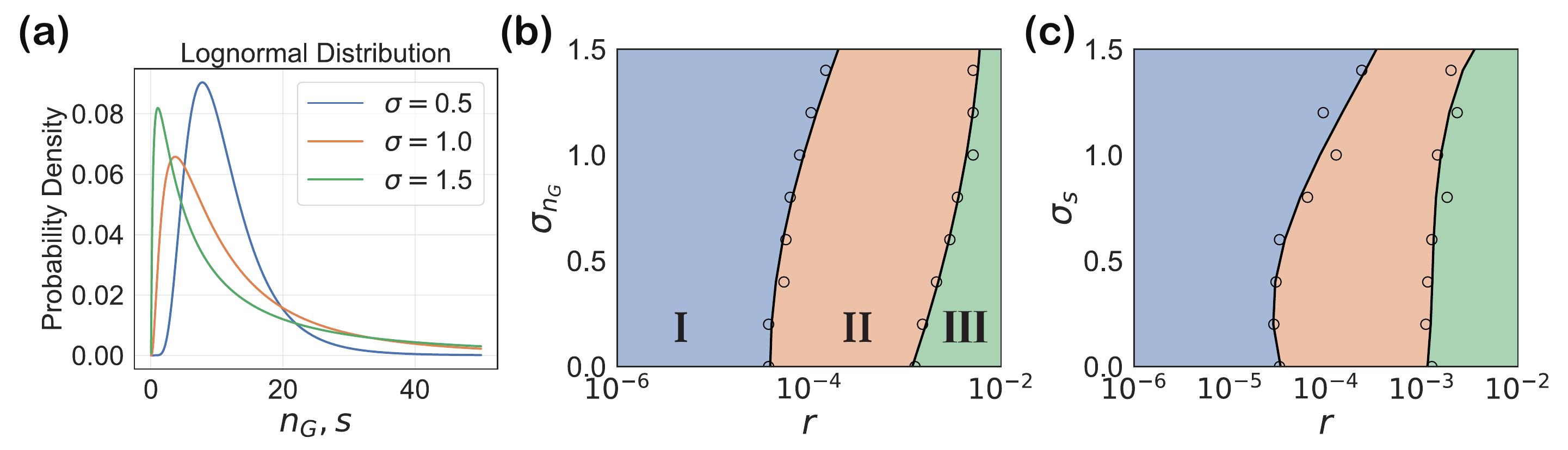}
\caption{Phase diagrams for $s$ and $n_G$ sampled from the log-normal distribution. The scatter points are the minimal HGT rates to keep gene and genotype diversity invariant over a simulation period $T=250000$. The black solid lines are spline interpolation through the scatter points.  $\sigma_n=0$ and $\sigma_s=0$ correspond to the simplified case shown in Fig. \ref{fig:phase} in the main text.
}
\label{figS:heter}
\end{figure}

Fig. \ref{figS:heter} shows that the results derived from our simplified model with identical phage-bacteria pairs are robust to heterogeneity and work pretty well for $\sigma<0.5$.

\section{Beyond one-bacteria-one-phage infection}\label{sec:mutltiple_strain_infection}
\begin{figure}[H]
\centering
\includegraphics[width=0.8\textwidth]{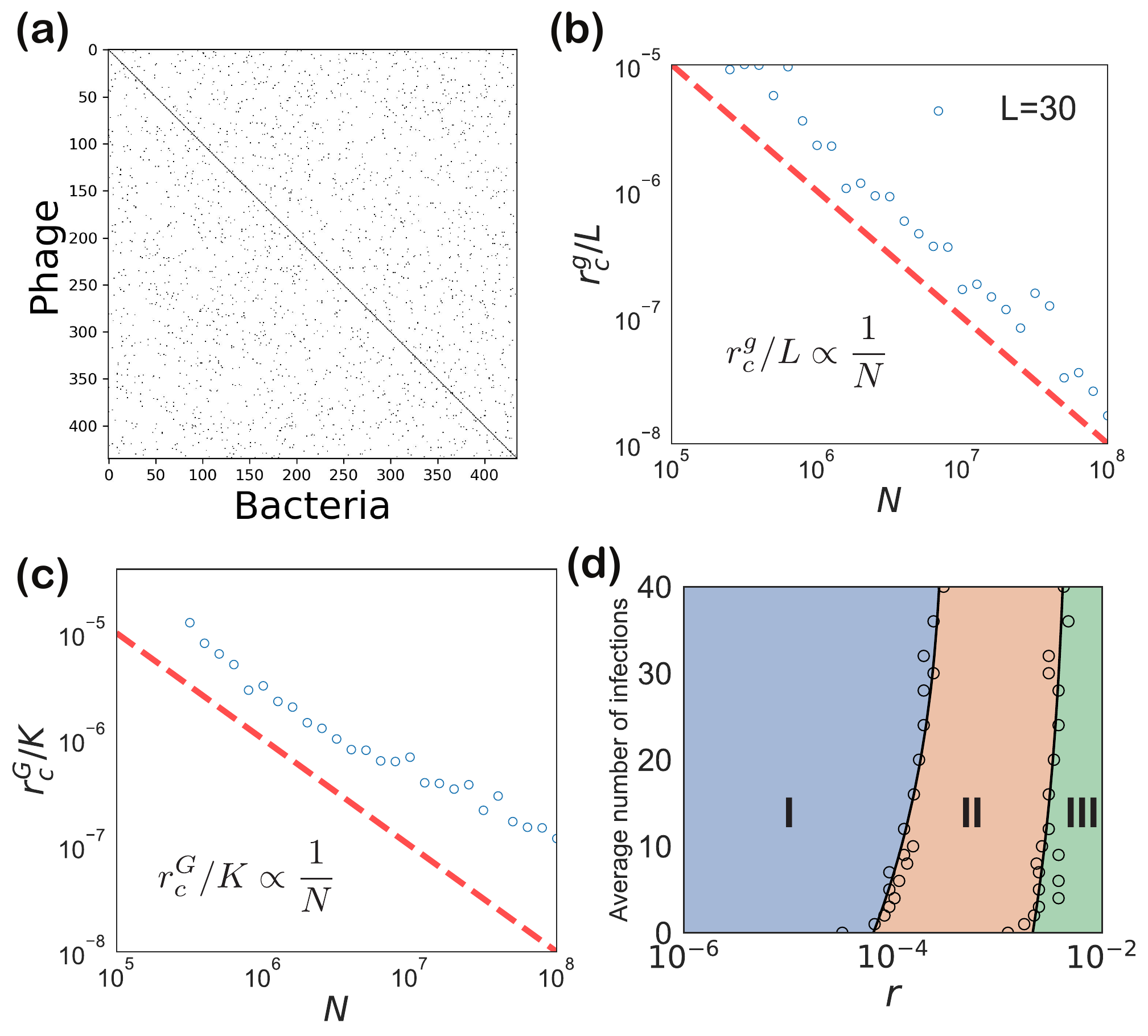}
\caption{{\bf(a)} The heatmap for the binary sparse interaction matrix with $L=40$. Each phage strain has the ability to infect another several random bacterial strains beside the exact matching one. The number of infections is sampled from the Poisson distribution with rate 5.  The $y/x$-axis are the labels of phages/bacteria. The black points are the non-zero entries, indicating the phage can infect that bacterium. {\bf(b, c)} Minimal HGT rates for gene and genotype coexistence  for inter-bacteria and phage-bacteria HGT. The red dashed line is our theoretical prediction, which shows our theory can also be generalized to the sparse interaction structure. {\bf(d)} shows how the phase diagram is distorted by varying the average number of infections, i.e., the Poisson rate. Here we fix $N=10^6$ and $L=40$.}
\label{figS:sparse_interaction}
\end{figure}

We extended our model to handle multiple phage-host infections by initializing a diagonal binary infection matrix first. We then generalized the interaction to allow a single phage strain to simultaneously infect multiple hosts, with the number of infections drawn from a Poisson distribution with a Poisson rate of 5, as illustrated in Fig. \ref{figS:sparse_interaction} Panel (a). Fig. \ref{figS:sparse_interaction} 
 Panels (b) and (c) demonstrate that our scaling relations remain consistent. Furthermore, Fig. \ref{figS:sparse_interaction}  Panel (d) illustrates the robustness of our results across a wide range of average infection numbers by systematically varying the Poisson rate.

\section{Transformation between the time and population distributions}\label{app:transformation}
In the bacterial booming process, the bacterial abundance grows exponentially with time:
\begin{equation}\label{eq:risingdynamics}
 n(t) = n_0e^{st}, \quad  d n = s n d t,
\end{equation}
where $s$ is its fitness, and $n_0$ is the initial population size at $t=0$.

We would like to estimate the population distribution from the time distribution,
\begin{equation}
\begin{aligned}
 Q(n) =\int P(n|t)P(t) dt =\int \delta(n_0e^{st} - n)P(t)dt =\int \delta(t-\frac{1}{s}\log\frac{n}{n_0})P(t) \frac{dt}{sn}= \frac{1}{sn}P(\frac{1}{s}\log{\frac{n}{n_0}}).
 \end{aligned}
\end{equation}

Reverse this process, and recall the normalization: $\int P(t|n) dt =1$, 
\begin{equation}
\begin{aligned}
 P(t) =& \int P(t|n)Q(n) dn =\int sn_0e^{st}\delta(n_0e^{st} - n)Q(n) dn =sn_0e^{st}  Q(n_0e^{st} ).
 \end{aligned}
\end{equation}

\section{Establishment probability with a constant fitness for two different processes}
\subsection{Birth-death process}
When the phage population is small, its targeted bacterium's fitness is $s$. In a unit of time $\Delta t$, the bacterium's birth probability is $(1+s)\Delta t$, its death probability  $\Delta t$, and the probability that neither happens is $1- (s+2)\Delta t $.  With the basic extinction theory of branching
processes,  its extinction probability after $t$ generations, $w(t)$, follows
\begin{equation}\label{eq:birth-death}
 w(t) = (1- (s+2)\Delta t) w(t-1) + \Delta t + (1+s)\Delta t w(t-1)^2,
\end{equation} 
where we assume each birth event happens with two offspring.  
In the continuum limit,
\begin{equation}
 \frac{dw}{dt}= 1 - (s+2)w  + (1+s) w^2.
\end{equation}
It is useful to rewrite in terms of the survival probability $\phi(t) = 1 - w(t)$:
\begin{equation}
 \frac{d\phi}{dt}= s\phi -(1+s)\phi^2.
\end{equation}
In the long-time limit,  it gives the establishment probability
\begin{equation}\label{eq:birthdeathphi}
 \phi(\infty) = \frac{s}{1+s}\approx s.
\end{equation}

\subsection{Poisson process}\label{AP:correctPoisson}
In our simulation, we use the Poisson process instead of the birth-death process with two offspring. Instead of eq. (\ref{eq:birth-death}), the extinction probability after time $t$  follows
\begin{equation}
 w(t) = e^{-(s+1)\Delta t}\sum^{\infty}_{k=0}\frac{((s+1)\Delta t )^k }{k!} w(t-\Delta t)^k= e^{(s+1)\Delta t [w(t-\Delta t) - 1]},
\end{equation}
where $e^{-(s+1)\Delta t}\frac{(s\Delta t )^k }{k!}$ is the probability of having $k$ offspring during a unit of time $\Delta t$, and $w(t-\Delta t)^k$ is the probability that all $k$ offspring go extinct after $t-\Delta t$. 

The expression in terms of the survival probability is
\begin{equation}
1 -\phi(t)  = e^{-(1+s)\Delta t \phi(t-\Delta t) }.
\end{equation}
Assuming $\phi(\infty)$ is small, and choosing $\Delta t=1$ in the simulation, the establishment probability is
\begin{equation}
\phi(\infty)  = \frac{2s}{1+2s}\approx 2s.
\end{equation}
Comparing with eq. (\ref{eq:birthdeathphi}), our simulation set-up brings a factor of 2.  We use the birth-death process in the weak HGT regime because it is easy to construct the partial differential equation (PDE) in the continuum limit. When comparing with the simulation, we will correct our theoretical results about the establishment probability by a factor of 2.

\section{Other supplementary figures}

\begin{figure}[H]
\centering
\includegraphics[width=0.9\textwidth]{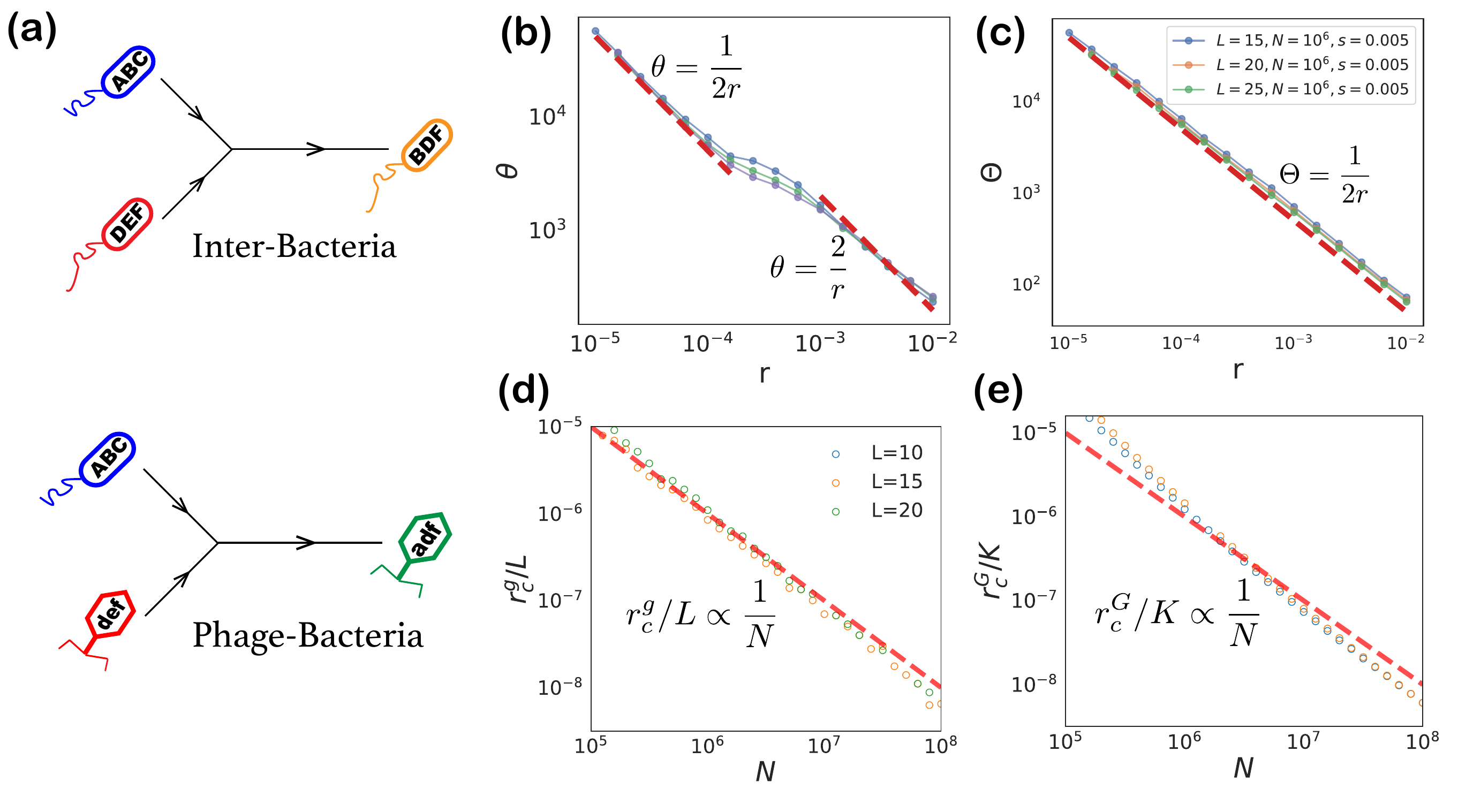}
\caption{{\bf(a)} Scheme of inter-bacteria and phage-bacteria HGT for "tripleton". The phage and bacterium can carry three different genes. There is one gene transferred in each HGT event.  {\bf(b)}  Effective genotype temperature $\Theta$ and {\bf(c)} gene temperature $\theta$ for "tripleton". Critical HGT rates for {\bf(d)} gene coexistence and {\bf(e)} genotype coexistence for the "tripleton" case. The red dashed lines are our theoretical predictions (the same as "doubleton" case).}
\label{figS:tripleton}
\end{figure}

\begin{figure}
\centering
\includegraphics[width=0.8\textwidth]{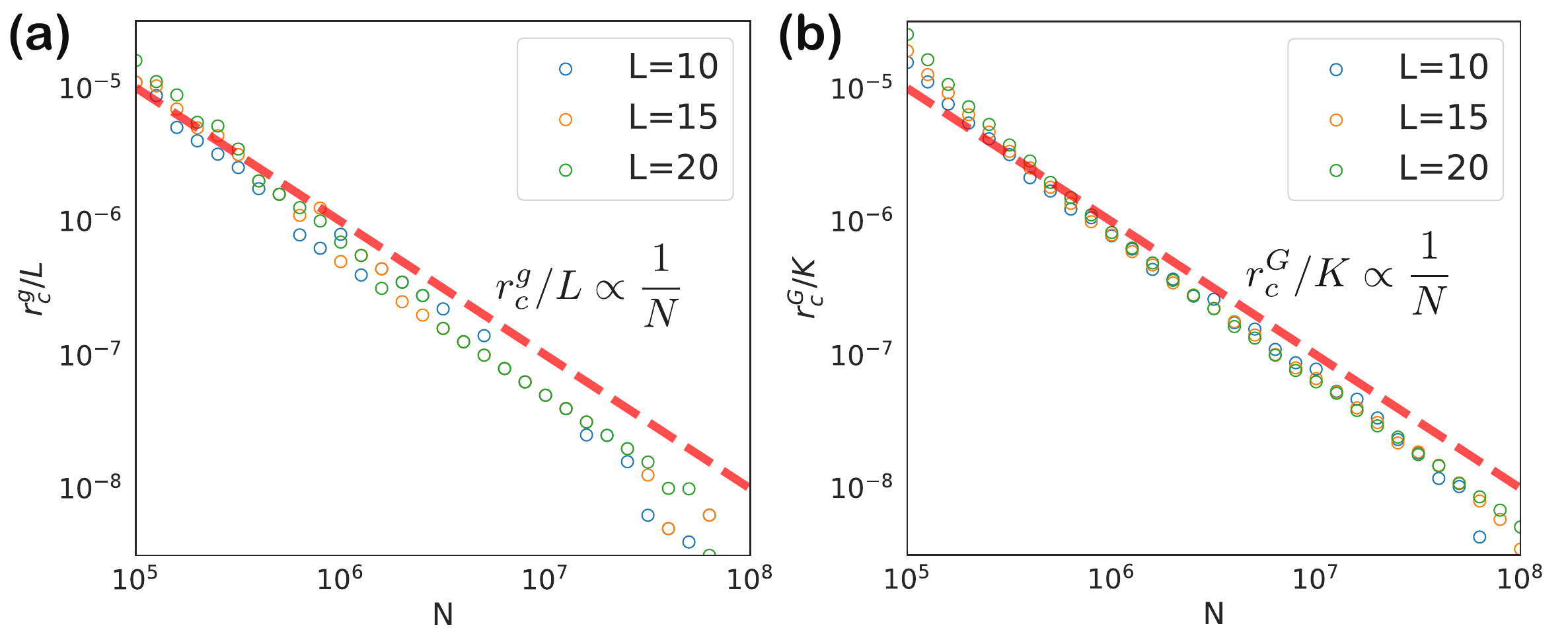}
\caption{{Minimal HGT rates for gene coexistence {\bf(a)}} and genotype coexistence {\bf(b)} for inter-bacteria and phage-bacteria HGT without fixing the total population size.  The red dashed line is our theoretical prediction, which shows the constraint of fixing the total population size does not affect our results in the main text.}
\label{figS:population_constraint}
\end{figure}

\end{document}